\documentclass[superscriptaddress,floatfix,aps,pre,twocolumn,notitlepage,shortbibliography,nofootinbib]{revtex4-2}

\usepackage[utf8]{inputenc}
\usepackage{amsmath}
\usepackage{amssymb}
\usepackage{refcount}
\usepackage{siunitx}
\usepackage{graphicx}
\usepackage{hyperref}
\hypersetup{colorlinks,allcolors=black}

\usepackage{natbib}
 \usepackage[capitalise]{cleveref}
\usepackage{microtype}
\usepackage{bm}
\usepackage{lipsum}
\usepackage[english]{babel}
\usepackage{color}
\usepackage{graphicx}
\usepackage{xcolor}
\usepackage{braket}

\newcommand{\comm}[2]{\left[ #1 , #2 \right]}
\newcommand{\acomm}[2]{\left\{ #1, #2 \right\}}
\newcommand{\abs}[1]{\left| #1 \right|}

\newcommand{\bp}{\mathbf{p}}

\begin{document}
\author{Haidar Al-Naseri}
\email{haidar.al-naseri@umu.se}
\affiliation{Department of Physics, Ume{\aa} University, SE--901 87 Ume{\aa}, Sweden}


\author{Gert Brodin}
\email{gert.brodin@umu.se}
\affiliation{Department of Physics, Ume{\aa} University, SE--901 87 Ume{\aa}, Sweden}
\title{Applicability of the Klein-Gordon equation for pair production in vacuum and plasma}

\begin{abstract}
 In this paper, a phase-space description of electron-positron pair-creation will be applied, based on a Wigner transformation of the Klein-Gordon equation. The resulting theory is similar in many respects to the equations from the Dirac-Heisenberg-Wigner formalism. However, in the former case, all physics related to particle spin is neglected. In the present paper we compare the pair-production rate in vacuum and plasmas, with and without spin effects, in order to evaluate the accuracy and applicability of the spinless approximation. It is found that for modest frequencies of the electromagnetic field, the pair production rate of the Klein-Gordon theory is a good approximation to the Dirac theory, provided the matter density is small enough for Pauli blocking to be neglected, and a factor of two related to the difference in the vacuum energy density is compensated for.     
   
\end{abstract}
 
\maketitle

\section{Introduction}
The interest in strong field physics has increased in the past decades due to the rapid development of the laser facilities \cite{ELI,SEL}.
See, e.g., Ref. \cite{Review1,Anton} for the recent developments in the research area.
In the presence of a strong electromagnetic field, the vacuum can decay into electron-positron pairs. This process was first studied by Sauter in 1931 \cite{Sauter} and a rigorous mathematical description of this process was derived by Julian Schwinger in 1951 \cite{Schwinger}.
Since then, the basic mechanism has been studied in a dynamical context and for more complicated field geometries. In particular, a study of the interplay between temporal and spatial variations of the fields has been done in Refs.\cite{Gies2,Gies1,Kohlfurst_spat1,Kohlfurst_spat2}. To maximize the number of produced particles in the case of sub-critical field strengths, a geometry of colliding laser pulses has been suggested in \cite{Colliding2,Colliding1,Colliding4,colliding3}.


In spite of much progress in recent years \cite{Gies1,Gies2,Kohlfurst_spat1,Kohlfurst_spat2,Colliding1,Colliding2,colliding3,Colliding4}, due to spin and chiral effects associated with Dirac fermions, various simplifying assumptions are typically needed  when studying electron-positron pair  production. A way to relax some of these assumptions concerning e.g. the electromagnetic field geometry, is to study scalar QED, where the spin of the particles is ignored. Scalar QED models, based on the Klein-Gordon equation, are considerably less complicated than those derived from the Dirac equation. While it is known that many qualitative features regarding pair-production are preserved when replacing fermionic QED with scalar QED \cite{ScalarQED7,Kohlfurst2,Kohlfurst1}, a better understanding of the differences/similarities are desirable, to know to what degree results from scalar QED can be trusted. The purpose of the present paper is to study pair production in vacuum and plasmas, with and without including the particle spin properties, to highlight differences and similarities between the bosonic and fermionic results.

 Various models and methods can be used when studying electron-positron pair production from vacuum and in plasmas.  This includes quantum kinetic theories (QKT), which can be applied for fermionic and bosonic pair production in electric fields \cite{Bloch}. Treatments, where a bosonic model of pair production has been used, are given in Refs. \cite{ScalarQED1,ScalarQED3}, and works on both bosonic and fermionic pair production in electric and magnetic fields can be found in Ref. \cite{ScalarQED4}. Moreover, lattice Quantum Electrodynamics has been used to model scalar QED pair production \cite{ScalarQED2}.
 
In the present paper, we will use a phase-space approach, based on a gauge invariant Wigner transform of the evolution equations. In the fermionic case, this leads to the real-time Dirac-Heisenberg-Wigner (DHW) formalism, first derived in Ref. \cite{Birula}. For the bosonic case, this corresponds to making a (gauge-invariant) Wigner transform of the Klein-Gordon equation, as was first derived in Ref. \cite{Best} with applications presented e.g. in Refs. \cite{Best,KG_wigner_Review,KG_wigner,KG_wigner_hydrodyn,ScalarQED5,ScalarQED6}. We will refer to these equations as the Klein-Gordon-Wigner (KGW) equations. We note that since both KGW and DHW are based on a phase-space approach, they are conceptually similar, and this facilitates a direct comparison of the results. 

An important conclusion from our study, which agrees with previous findings \cite{Schwinger,Nikishov}, is that KGW and DHW are in exact agreement for pair-production in the zero frequency limit, provided a factor of two is compensated for. The factor of two is closely related to the difference between the fermionic and bosonic vacuum expectation values. When the frequency of the applied field is increased, the spin polarization current in the DHW-formalism grows, which is associated with a gradually larger deviation between the bosonic and fermionic results. For significant pair-production, the self-consistent field generated by the plasma currents becomes appreciable, and this tends to enhance the deviation between bosonic and fermionic dynamics. For even higher plasma densities, Pauli blocking can be important, which naturally cannot occur in the bosonic case.     



The paper is organized as follows: In \cref{Section1} we present the derivation of the KGW-formalism. In \cref{Section2} we derive the linear response of a plasma to electromagnetic waves. We make a comparison both to the classical dispersion relation (which is recovered in the appropriate limit) and to the results based on the DHW formalism.
Next, numerical results for large fields (Schwinger) pair production are presented in \cref{Section3}. Here we compared the particle production in KGW and DHW using different scaling parameters. Finally, we present our conclusions in \cref{Section4}.

\section{KGW-model}
\label{Section1}
In this section, we will make a brief derivation of the Klein-Gordon-Wigner (KGW) formalism that has previously been derived by Ref. \cite{Best}, which the reader may consult for more details. The starting point is the Klein-Gordon equation
\begin{equation}
    \Big[ (\partial_{\mu}-ieA_{\mu})(\partial_{\mu}+ieA_{\mu}) + m^2 \Big] \phi(\mathbf{r},t)=0
\end{equation}
where $e$ is the elementary charge, $m$ is the mass of the electron, and $A^{\mu}=(A^0,\mathbf{A})$, where $A^0$ $\mathbf{A}$ are the scalar vector potential, respectively.
This equation contains a second-order derivative in time. To transform the Klein-Gordon equation to phase-space and obtain an explicit expression of the charge density in phase-space, we express the Klein-Gordon equation in the representation of Feshbach and Villars \cite{Feshbach}. In this representation, we have a first-order time-derivative. The Klein-Gordon field is expressed by the two-component wave function
\begin{equation}
\Phi =   
\begin{pmatrix}
\psi \\
\chi
\end{pmatrix}
\end{equation}
where 
\begin{align}
    \psi&=\frac{1}{2} \Big( \phi +\frac{i}{m}\frac{\partial \phi}{\partial t}-\frac{eA^0}{m}\Big)\\
    \chi&=\frac{1}{2} \Big( \phi -\frac{i}{m}\frac{\partial \phi}{\partial t}+\frac{eA^0}{m}\Big)
\end{align}
In matrix representation we get
\begin{multline}
\label{Klein2}
    i\frac{\partial \Phi}{\partial t}=\Bigg[ \frac{1}{2m}\Big(-i\frac{\partial}{\partial \mathbf{r}} -e\mathbf{A}\Big)^2
    \begin{pmatrix}
    1&1\\
    -1&-1
    \end{pmatrix}
    \\
    +m
        \begin{pmatrix}
    1&0\\
    0&-1
    \end{pmatrix}
    +eA^0 \mathbf{1}
    \Bigg] 
    \Phi
\end{multline}
where $\mathbf{1}$ is the identity matrix. This equation is first order in the time-derivative and has a Schrödinger-type of time evolution. The right-hand side of this equation can be interpreted as a Hamiltonian operator that will be used to find the time evolution in phase-space. Next, we use the gauge-invariant Wigner transformation
\begin{multline}
\label{Wigner}
    \hat{W}(\mathbf{r},\mathbf{p},t)\\=\int d^3z \exp
    \bigg(i \mathbf{p}\cdot \mathbf{z}+ie \int ^{1/2}_{-1/2} d\lambda \mathbf{z}\cdot \mathbf{A}(\mathbf{r}+\lambda \mathbf{z},t)\bigg)
    \hat{C}(\mathbf{r},t)
\end{multline}
where 
\begin{equation}
    \hat{C}(\mathbf{r},t)=\acomm{\Phi (\mathbf{r}+z/2,t)}{ \Phi (\mathbf{r}-z/2,t)}
\end{equation}
The Wigner function is defined as the expectation value of the Wigner operator
\begin{equation}
    W(\mathbf{r},\mathbf{p},t)=\bra{\Omega} \hat{W}(\mathbf{r},\mathbf{p},t) \ket{\Omega}
\end{equation}
where $\ket{\Omega}\bra{\Omega}$ is the state of the system in the Hilbert space. To find an evolution equation in phase space, we take the time derivative of \cref{Wigner} and use the Hamiltonian operator in \cref{Klein2}. We use the Hartree approximation where the electromagnetic field is treated as a non-quantized field. Finally, we have an equation of motion of the Wigner function
\begin{multline}
\label{Equation_of_motion}
    iD_t W(\mathbf{r},\mathbf{p},t)=-\frac{i}{2}\hat{O}_1 \acomm{\sigma_3+i\sigma_2}{W}\\+ \hat{O}_2\comm{\sigma_3+i\sigma_2}{W}
    +m\comm{\sigma_1}{W}
\end{multline}
where $\sigma_i$, with $i=1,2,3$, are the Pauli matrices and

\begin{align}
   \hat{ O}_1&=\frac{\mathbf{p}\cdot \nabla}{m}+\frac{\mathbf{p}}{m}\cdot e\int^{1/2}_{1/2}d\lambda \mathbf{B}(x+i\hbar \lambda \nabla_p)\times \nabla \\
    \hat{O}_2 &=\frac{\nabla^2}{4m}-\frac{p^2}{m}
    -\frac{e\hbar^2}{12m} \nabla\cdot (\mathbf{B}\times \nabla)\notag \\
    &+2\frac{\mathbf{p}}{m}\cdot ie\hbar 
    \int^{1/2}_{1/2}d\lambda \lambda \mathbf{B}(x+i\hbar \lambda \nabla_p)\times \nabla_p\\
    D_t&=\partial_t+ e\int_{1/2}^{1/2}d\lambda \mathbf{E}(x+i\hbar \lambda \nabla_p)\cdot\nabla_p
\end{align}
For longer macroscopic scales, the non-local operators given by the integrals can be expanded in powers of $\hbar$. Thus, for example in the long-scale limit, we may use $D_t=\partial_t+e \mathbf{E}\cdot \nabla_p$, and similarly for the other operators \cite{Best}.
The interpretation of the variables described by \cref{Equation_of_motion} is not simple. Thus we make an expansion of the Wigner function over the Pauli matrices $\sigma_i$ and the identity matrix $\mathbf{1}$
\begin{equation}
    W(\mathbf{r},\mathbf{p},t)= f_3(\mathbf{r},\mathbf{p},t)\mathbf{1}+ \sum_{i=1}^{3}f_{3-i}(\mathbf{r},\mathbf{p},t) \sigma_i
\end{equation}
where the expansion coefficients $f_i$, with i=0-3, will lead to four coupled partial differential equations.

\begin{align}
\label{PDE-system}
    D_tf_0&=-\hat{O}_1 (f_2+f_3)\notag \\
    D_tf_1&=-\hat{O}_2(f_2+f_3)+2mf_2 \notag \\
    D_tf_2&= \hat{O}_2 f_1 + \hat{O}_1f_0 -2mf_1\notag \\
    D_tf_3&=-\hat{O}_2f_1 -\hat{O}_1f_0
\end{align}
This PDE-system is the final one that covers all physics of the Klein-Gordon equation. 
This system is closed by Maxwell's equations with the sources

\begin{align}
     \rho&=e\int d^3pf_0\\
    {\bf J}&=\frac{e}{m} \int d^3p\, \mathbf{p}(f_2+f_3)
\end{align}

Compared to the DHW-formalism \cite{Birula} which is based on the Dirac equation, this system contains only four instead of 16 scalar equations. This makes the KGW-formalism simpler to solve numerically. 
The 4 scalars of the KGW-formalism have a straightforward physical interpretation, as can be seen from the following observables
\begin{align}
    Q&= e\int d^3pd^3x f_0\\
    \mathbf{J}&= \frac{e}{m} \int d^3p\, \mathbf{p}(f_2+f_3)\\
    W&= \int d^3pd^3x \Big[ \frac{p^2}{2m}(f_2+f_3)+mf_3\Big]\\
    \mathbf{M}&= \int d^3pd^3x \, \mathbf{p} (f_0-f_1)
\end{align}
where $Q$ is the total charge, $\mathbf{J}$ is the current density, $W$ is the particle energy and $\mathbf{M}$ is the momentum. Interpretations that can be done from the above expressions include, e.g.,  that $e f_0$ is the (phase space) charge density and $e\mathbf{p} /m(f_2+f_3)$ the current density, as implied by the sources in Maxwell's equations. 

The $f_i$-functions have the vacuum-contribution
\begin{align}
\label{vacuum}
    f_0=f_1=0 \notag \\
    f_2+f_3=\frac{m}{\epsilon}\notag \\
    f_3-f_2=\frac{\epsilon}{m}
\end{align}
where $\epsilon=\sqrt{m^2+p^2}$. The expressions above are obtained by calculating the Wigner operator for the free particle Klein-Gordon equation and taking the vacuum expectation value. If we add a plasma to the background, we should modify the source terms for $f_2+f_3$ and $f_3-f_2$. Moreover, the charge density $f_0$ should be non-zero. Adding electron and positron sources we first obtain $f_2+f_3=(m/\epsilon)F$ and $f_3-f_2=(\epsilon/m)F$ in \cref{vacuum} where
\begin{equation}
\label{KGW_F}
 F=1+2f_e(\mathbf{p}) + 2f_p(\mathbf{p})
\end{equation}
Here $f_{e,p}$ can be viewed as classical electron/positron distribution functions.
Note that we have a positive sign between the particle and vacuum sources, instead of a negative one as in the DHW-formalism \cite{Birula,2021}. This is related to the physics of the Pauli-exclusion principle, which obviously is not included in the KGW-formalism, as the model is derived for spinless particles. By contrast, in the DHW-formalism, the Pauli-principle will be respected provided the correct expectation values for the vacuum states are implemented. Moreover, we can note that the particle contributions to $F$ have an extra factor of $2$ in \cref{KGW_F}, as compared to DHW-formalism. This is due to the fact that the magnitude of the vacuum contribution is only half as large for spinless particles.
The functions $f_{e/p}$ can be picked as any common background
distribution function from standard kinetic theory, i.e., a
Maxwell-Boltzmann, Synge-Juttner, or Fermi-Dirac distribution, depending on whether the characteristic kinetic energy is
relativistic and whether the particles are degenerate.
In conclusion, initial conditions for the KGW-variables involving a homogeneous medium with electron and positron distribution functions $f_{e,p}$ are given by
\begin{align}
\label{Initial_total}
    f_0&=2f_p-2f_e \notag \\
    f_1&=0 \notag \\
    f_2+f_3&=\frac{m}{\epsilon}F\notag \\
    f_3-f_2&=\frac{\epsilon}{m}F
\end{align}
with $F$ given by \cref{KGW_F}. The above expression gives a time-independent solution to the KGW-equations in the absence of electromagnetic fields (assuming that the total charge density and current density are zero). The expressions in \cref{Initial_total}, describing the equilibrium states, are a necessary prerequisite for including dynamical scenarios, leading to particle pair creation as will be studied in the next sections.

\section{Linear waves}
\label{Section2}
In this section, we consider the linear response of a plasma to an electromagnetic wave field, accounting also for the contribution from the vacuum expectation values.
For our case with no background electromagnetic fields, as described in the previous section, we get the
unperturbed vacuum contributions as the Wigner transform of
the expectation value of the free Klein-Gordon field operators. Thus, \cref{Initial_total} (including the plasma) applies for the background.
For the wave perturbation, we consider electromagnetic waves with the following geometry
\begin{align}
    \mathbf{E}&=E \mathbf{e}_x\notag \\
    \mathbf{k}&=k \mathbf{e}_z\notag \\
    \mathbf{B}&= B \mathbf{e}_y
\end{align}
The ions will not be treated dynamically but will constitute a constant neutralizing background. In the absence of ions, naturally, the electron and positron background charge densities must cancel for the background to be in equilibrium. However, in the presence of a constant (immobile) ion background, the electron and positron may differ, in particular, the positron background may vanish. 

Next, in order to study linear theory, we divide the functions $f_i$ as
\begin{equation}
    f_i(z,\bp,t)=f_i^0(\bp)+f_i^1(\bp)e^{i(kz-\omega t)}
\end{equation}
where the upper indices denote unperturbed and perturbed quantities. Furthermore, we use Ampére's law to obtain an implicit dispersion relation that can be written as
 \begin{multline}
     D(k,\omega)= \omega^2-k^2+ i\omega J_x\\=\omega^2-k^2+\frac{ie\omega }{(2\pi\hbar)^3E}\int d^3p\frac{p_x}{m}(f_3^1+f_2^1)=0
 \end{multline}
After some algebra, we can compute the current sources and deduce the explicit expression
\begin{widetext}

\begin{multline}
\label{KHW}
    D(k,\omega)=\omega^2-k^2 + \frac{ e^2}{(2\pi \hbar )^3}\int d^3p \frac{p_x}{\omega^2-k^2-4\epsilon^2+4k^2p_z^2/\omega^2}\Bigg[
    (\omega^2-2(p+\hbar k/2)^2-4m^2)\times \\
    \Big[\Delta_1 \nabla_{p_x}
    - \frac{2\hbar k}{\omega^2}\Big(\frac{\hbar \mathbf{k}}{12}-2\mathbf{p}\Delta_2 \Big)\cdot (\mathbf{e}_y\times \nabla_p) \Big] \Big( \frac{2m^2+ p^2}{2m\epsilon}F -\frac{ p^2}{2m\epsilon}F  \Big)
    -2\Delta_1 \nabla_{p_x} \frac{p^2}{\epsilon}F
    +8\frac{k}{\omega} p_x \Delta_1 \nabla_{p_z} (f_e-f_p)\\
    -4\frac{k^2}{\omega^2}p_z(p_x\nabla_{p_z}-p_z\nabla_{p_x})\Delta_1 \frac{F}{\epsilon}
    -4\Big((p+\hbar k/2)^2+2m^2\Big )\frac{\hbar k}{\omega^2}\Big(\frac{\hbar \mathbf{k}}{12}-2\mathbf{p}\Delta_2 \Big)\cdot (\mathbf{e}_y\times \nabla_p) \frac{F}{\epsilon}
    \Bigg]=0
\end{multline}
 \end{widetext}
 where 
\begin{align}
     \Delta_1&=\int^{1/2}_{1/2}d\lambda \cos{(\hbar k \lambda \nabla_{pz})}\\
     \Delta_2&= \int^{1/2}_{1/2}d\lambda \lambda \sin{(\hbar k \lambda \nabla_{pz})} 
 \end{align}
Here the operators are given by $\Delta_1=1$ and $\Delta_2= \hbar k\nabla_{pz}/12$ in the long-scale limit, where we have kept contributions to the lowest non-vanishing order.
We can obtain the classical limit by letting $\hbar \longrightarrow 0$ in \cref{KHW}. Doing so, we get $\Delta_2=0$. The dispersion relation then reduces to 
 \begin{multline}
     D(k,\omega)=\omega^2-k^2 -\frac{e^2}{\hbar^3 \pi^2} \int dp\, p^2
     \bigg[\Big(1-\frac{p^2}{3\epsilon^2} \Big)\frac{f_{e}+f_p}{\epsilon}\\
     -\frac{kp^2}{3\epsilon^2}\Big(\frac{1}{\omega+kp_z/\epsilon}
     -\frac{1}{\omega-kp_z/\epsilon}
     \Big)\frac{\partial (f_{e}+f_p) }{\partial p_z}
     \bigg]=0
 \end{multline}
which can be shown to agree with the standard result based on the relativistic Vlasov equation, after
some straightforward algebra. 
Here, $f_e$ and $f_p$ behave as classical distribution functions with one important difference. Since, in the underlying model, electrons and positrons are described by the same evolution equations, we must model the case of two identical beams moving with the same beam momentum $p_d$ as $f_e(p-p_d)= f_p(p+p_d)$.
This accounts for the fact that positrons can be viewed as electrons moving backward in time.
We note that the appearance of $\hbar$ in the integration measure is just a matter of normalization, and not a sign of any remaining quantum features \cite{RMPP-review}.
 
Now, we want to compare the dispersion relation \cref{KHW} with the DHW formalism. To be able to do that, we take the homogeneous limit of \cref{KHW} (since only the electrostatic limit resulting from $k\longrightarrow 0$ has been computed in the DHW-case) and compare it with the corresponding result from the DHW-formalism \cite{2021,2022linear}.
 Letting $k\longrightarrow 0$ in \cref{KHW}, and including electrons only in the background plasma $f=f_e$, we get
 \begin{multline}
 \label{Disp_hom_KGW}
\omega^2=-\frac{4e^2}{\pi^2 \hbar^3} \int dp \frac{p^2 \epsilon}{\hbar ^2\omega^2-4\epsilon^2} \times  \\
     \bigg[ \Big( 1-\frac{p^2}{3\epsilon^2}  -\frac{\hbar^2\omega^2}{4\epsilon^2}  \Big)f 
    - \frac{\hbar^2\omega^2 p^2}{24\epsilon^4} 
     \bigg]
 \end{multline}
 For the DHW formalism, we apply the $k\longrightarrow 0$- result of the electrostatic case, see \cite{2021,2022linear}, and we obtain
\begin{multline}
 \label{Disp_hom_DHW}
   \omega^2=-\frac{4e^2}{\pi^2\hbar^3 }\int dp \frac{p^2 \epsilon}{\hbar^2\omega^2-4\epsilon^2}\times \\
    \Big(1
    - \frac{p^2}{3\epsilon^2 } \bigg) 
    \bigg(
f-  \frac{\hbar^2\omega^2}{4\epsilon^2}
    \bigg) 
\end{multline}
We can clearly see that the results agree in the classical limit, and that the denominators coincide, i.e., the physics related to wave-particle interaction is similar.
Both for the KGW and DHW, the real part of the wave frequency $\omega_r$ is given by the classical limit to a good approximation.
The reason is that for $\hbar \omega \sim m$, the
Fermi energy $E_F$ will be much larger than unity. Thus, even if
$\hbar \omega\sim m$ we will have $\hbar \omega \ll E_f$, which, in turn, implies a minor quantum contribution to \cref{Disp_hom_KGW,Disp_hom_DHW}.
Nevertheless, there are also some differences between \cref{Disp_hom_KGW,Disp_hom_DHW}. 
To focus on the most important one, we now concentrate on the damping given by the imaginary part of the wave frequency $\omega_i$, associated with the pole contribution of the momentum integral.
Using the Plemelj formula

\begin{equation}
    \frac{1}{u-a}=P \frac{1}{u-1}+ i\pi \delta(u-a)
\end{equation}
where $P$ is the principal value, 
we obtain from the KGW-expression \cref{Disp_hom_KGW}
 \begin{equation}
  \label{ImD_KGW}
     \omega_i = -\frac{e^2 p_{res}^3}{6\hbar^4 \omega_r^4 \pi } \Big(1+ 2f(p_{res}\Big)
 \end{equation}
 For the DHW-system, we instead obtain 
 \begin{equation}
 \label{ImD_DHW}
\omega_i
=-\frac{e^2 p_{res}}{12\pi \hbar^2 \omega  } \Big(1+\frac{2m^2}{\hbar^2\omega_r^2}\Big)\Big(1-f(p_{res})\Big) 
\end{equation}
Here $p_{res}$ is the resonant momentum, making the denominators in the integrands zero, i.e. $m^2+p_{res}^2=\hbar^2\omega^2/4$. While some features of the two damping rates agree, it can be noted that there are also significant differences. These differences can be better understood in light of certain numerical results presented in the next section. Thus we will wait to make a more detailed comparison of the damping rates.  

\section{Numerical results}
\label{Section3}
\subsection{Preliminaries}
Before presenting the numerical results, we discuss the issue of renormalization.
For large momentum where we can use the approximation $\epsilon\approx p$, it is straightforward to see that the integrand in \cref{Disp_hom_KGW} scales as $1/p$, and thus the integral has a logarithmic divergence.
In principle, this needs to be dealt with, but in practice, for a numerical solution, this
is almost automatically solved by a numerical cut-off in the momentum integration that acts as an effective
regularization. Due to the slow growth of the logarithmic divergence, in the presence of a numerical momentum cut-off, simply ignoring the issue of renormalization does not result in a significant error. The technical aspects are the same as for the DHW-case, which has been described in more detail in Ref. \cite{2022linear,PRESchwinger}.



To able to compare the results of the numerical solution of \cref{PDE-system} with the ones from the DHW-formalism in \cite{PRESchwinger}, we consider the homogeneous-limit of \cref{PDE-system}
 \begin{align}
     D_t f_0&=0\notag \\
     D_tf_1&=\frac{p^2}{m} (f_2+f_3) + 2mf_2\notag \\
     D_t f_2&= -\Big(\frac{p^2}{m}+2m\Big) f_1\notag \\
     D_tf_3&=\frac{p^2}{m}f_1
 \end{align}
 Apparently, in the homogeneous limit, $f_0$ decouples, and we only need to solve three equations. To simplify the numerical solution, we define new variables as deviations from the vacuum state, i.e.
 \begin{equation}
     \Tilde{f}_i=f_i-f_{i,vac}
 \end{equation}
 where 
 \begin{align}
     f_{1,vac}&=0\notag \\
     f_{2,vac}&= \frac{1}{2m\epsilon}(m^2-\epsilon^2)\notag \\
 f_{3,vac}&= \frac{1}{2m\epsilon}(m^2+\epsilon^2)
\end{align}
With this choice, we make sure that our basic variables become small for large momenta, such that a momentum cut-off is possible. By introducing $\Tilde{f}_i$, we remove the vacuum contribution from the variables that we solve numerically. Obviously, this does not mean that vacuum physics is not included in our solution, we only do this to handle certain numerical technicalities.
The PDE-system is now
 \begin{align}
     D_t\Tilde{f}_1&=\frac{p^2}{m} (\Tilde{f}_2+\Tilde{f}_3) + 2m\Tilde{f}_2\notag \\
     D_t \Tilde{f}_2&= \frac{eEp_z}{2m\epsilon} \Big(1+\frac{m^2}{\epsilon^2}\Big) -\Big(\frac{p^2}{m}+2m\Big) \Tilde{f}_1\notag \\
     D_t\Tilde{f}_3&=-\frac{eEp_z}{2m\epsilon}p^2+\frac{p^2}{m}\Tilde{f}_1
 \end{align}
 For notational convenience, we omit the tilde for variables in what follows.
 In order to solve the system numerically, we need to make some further adaptions. The next step is to make the additional change of variable


 \begin{align}
     f_1&=f_1\notag \\
     f_{+}&=f_2+f_3\\
     f_{-}&=f_2-f_3\notag 
 \end{align}
introduce the canonical momentum $q$ given by
 \begin{equation}
     q=p+eA
 \end{equation}
and use normalized variables:
$t_n= \omega_{ce} t$, $q_n=q/m$, $p_{n\perp}=p_{\perp}/m$, $E_n=E/E_{cr}$, $A_n=eA/m$, where $\omega_{ce}$ is the Compton frequency.
The final system resulting from these changes, which will be solved numerically is
  \begin{align}
  \label{PDE-system_KGW}
     \partial_{t_n}f_1&=\epsilon_n^2 f_+ +f_- \notag \\
     \partial_{t_n}f_+&=\frac{(q_n-A_n) }{\epsilon_n^3}E_n-2f_1\\
     \partial_{t_n}f_-&=\frac{q_n-A_n}{\epsilon_n}E_n -2\epsilon_n^2 f_1 \notag 
 \end{align}

 together with Ampere's law
      \begin{equation}
\label{Ampers_law_KGW}
\frac{\partial E_n}{\partial t_n}=- \eta \int d^2p_n (q_n-A_n) f_{+}
\end{equation}
where $\eta=\alpha/\pi$, here $\alpha$ is the fine-stucture constant.
Note that, for notational convenience, we omit the subscript "n" for variables in what follows.
The system \cref{PDE-system_KGW}-\cref{Ampers_law_KGW}
is solved numerically using a phase-corrected staggered leapfrog method \cite{Leapfrog}.
 Typical parameters of the simulations are a time-step of the order of $\Delta t\sim 0.001$, a parallel momentum step $\Delta q\sim 0.01$, and a perpendicular momentum step $\Delta p_{\bot} \sim 0.1$. In spite of the rather good resolution in parallel momentum, the $q$-dependence of the produced data tends to look noisy in momentum space. This is due to the zitterbewegung effect which produces increasingly short scales. Nevertheless, the dynamics of the larger scales are not sensitive to the small-scale momentum details, i.e. changing $\Delta q$ does not affect the results presented in this paper.

 To confirm that the numerical scheme produces sound results, we study the energy conservation law. An energy conservation law of the system \cref{PDE-system_KGW,Ampers_law_KGW} can be written in the form 
\begin{equation}
    \frac{d}{dt}\left[ \frac{E^2}{2}+ \frac{\eta}{2}\int d^2p \Big(  (\epsilon^2+1) f_{+}  -f_{-}\Big) \right]=0
\end{equation}
 For the numerical resolutions used in the runs presented below, the total energy of the system is conserved  within a relative error typically less than $10^{-4}$.


\subsection{Effects due to the vacuum background}

\begin{figure}
    \centering
    \includegraphics[width=\linewidth, height=10 cm]{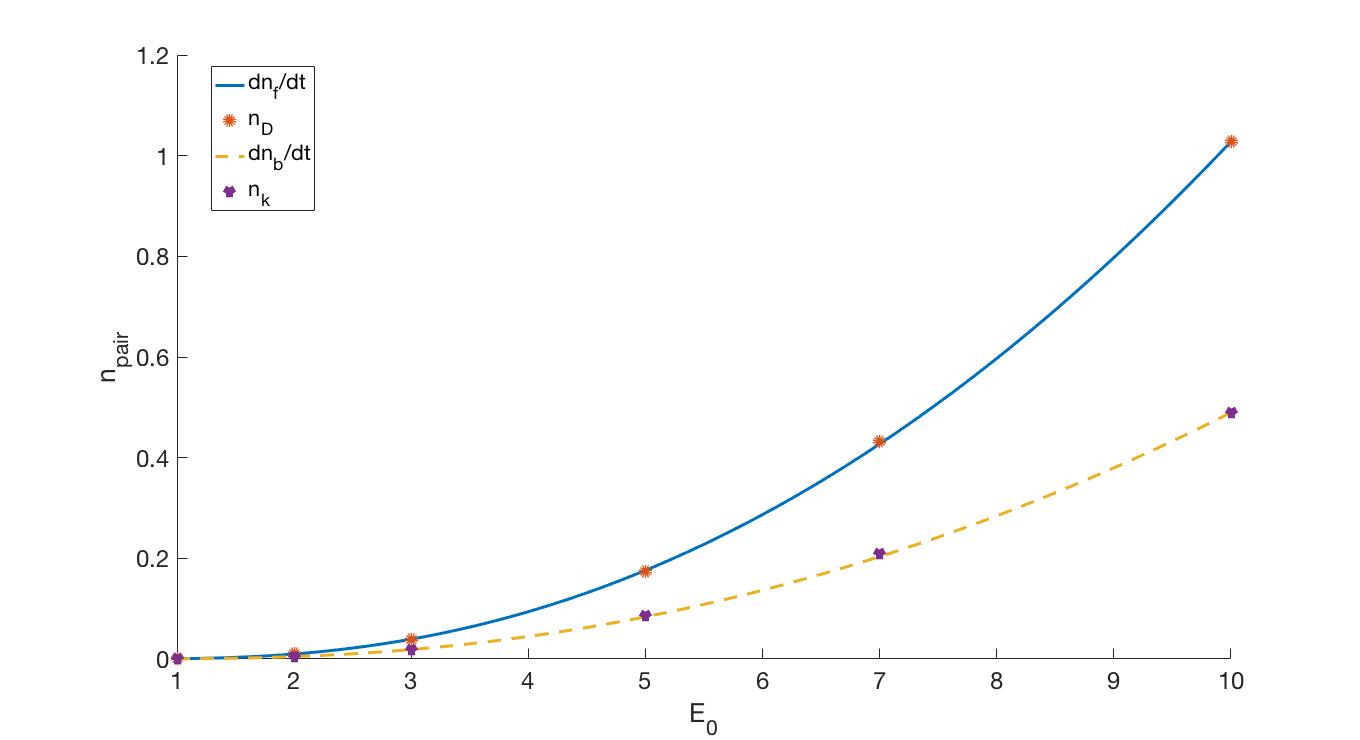}
    \caption{The number of produced pairs from KGW ($n_k$) and DHW ($n_D$) using the numerical solutions together with the corresponding analytical results $d n_b/dt$ and $d n_f/dt$. }
    \label{Constant_E}
\end{figure}
\begin{figure}
    \centering
    \includegraphics[width=\linewidth, height=10 cm]{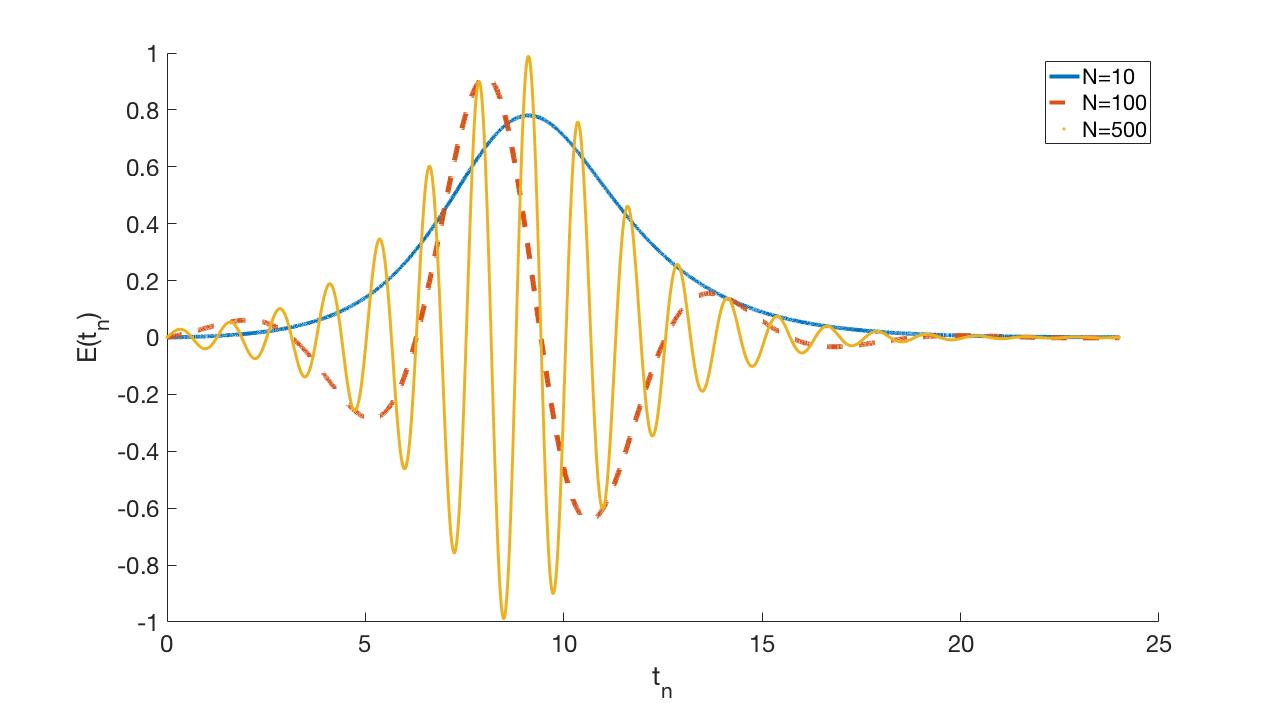}
    \caption{The electric pulse is plotted over time for three different frequencies $N=(10,100,500)$. }
    \label{Pulse_plot}
\end{figure}

Before solving the \cref{PDE-system_KGW}-\cref{Ampers_law_KGW} for more complicated scenarios, we can check the validity of the numerical solution by considering constant electric fields $E=E_0$. The rate for creating a fermionic pair from vacuum  $dn_f/dt$ is well-known \cite{Schwinger,Nikishov} 

\begin{equation}
\label{Schwinger_rate}
    \frac{dn_f}{dt}=\frac{1}{4\pi^3}\Big(\frac{eE}{\hbar}\Big)^2 \exp{\Big( - \frac{\pi E_{cr} }{ E}\Big)}
\end{equation}
By comparison, the rate for creating a bosonic pair as described by the KGW-theory, is $dn_b/dt= 0.5dn_f/dt$. This is a direct consequence of the factors given in \cref{KGW_F}, where the relative contribution from the vacuum sources is half of the contribution seen in the DHW-formalism.

To validate the numerics, we will solve
\cref{PDE-system_KGW} and the corresponding equations in the DHW-formalism \cite{2021} for the case of constant electric fields $E_0$ and only vacuum initially. Since our goal is to check the validity of the numerics by comparing with theory without back-reaction effects, we also neglect back-reaction, i.e., we don't solve \cref{Ampers_law_KGW} self-consistently. Then, we calculate the number of produced particles $n_k$ in the KGW-formalism and $n_D$ in the DHW-formalism, with $n_k$ given by
\begin{equation}
\label{Rate_KGW}
    n_k=\frac{1}{(2\hbar \pi)^3}\int d^3p \frac{1}{2\epsilon}\Big[(\epsilon+1) f_{+} - f_{-} 
    \Big]
\end{equation}
In \cref{Constant_E}, we compare the numerical values of $n_{k/D}$ with $dn_{b/f}/dt$ for different values of $E_0$. As can be seen, the numerical values follow the analytical ones with good agreement.

Interestingly, nothing prevents us from making a toy model where we use spinless dynamics for the evolution equations, but increase the vacuum expectation values by a factor of two in the KGW-model. Doing this, for constant fields $n_k$ will get the same behavior as $n_D$ and thus the slightly modified version of the KGW-formalism gives the exact same results as the DHW-formalism 

\subsection{High frequency effects}
Now, we turn to the problem of time-dependent electric fields. We
solve \cref{PDE-system_KGW}-\cref{Ampers_law_KGW} for the case with no plasma and only vacuum initially using the following representation of the time-dependent pulse
\begin{equation}
\label{Pulse}
    E(t)=E_0 {\rm sech}\Big(\frac{t}{b} -\tau_{0}\Big)\sin{\omega t}
\end{equation}
where $\tau_0$ is the phase-shift, $b$ is the pulse length and $\omega=N \omega_{ce}/100$. See \cref{Pulse_plot} for a temporal plot of the electric pulse.
We use $n_k$ \cref{Rate_KGW} to find the ratio $n_k/n_D$ of the produced pair in KGW-formalism and $n_D$ in the DHW-formalism  \cite{2021}. For constant fields $E=E_0$, given by the rate presented for fermions in \cref{Schwinger_rate} and for bosons, we have
\begin{equation}
    \frac{n_k}{n_D}=\frac{1}{2}
\end{equation}
\begin{figure}
    \centering
    \includegraphics[width= \linewidth, height=10 cm]{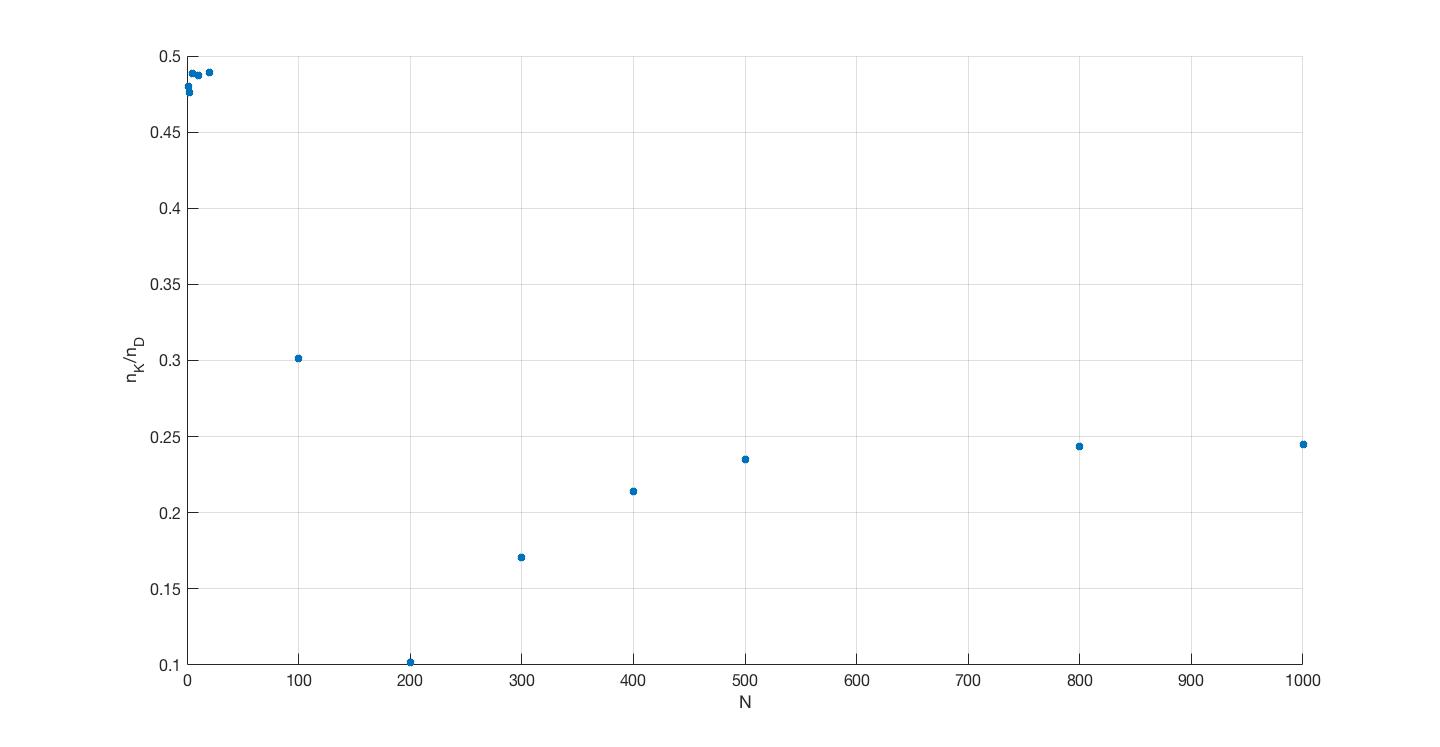}
    \caption{The ratio $n_k/n_D$ is plotted as a function of $N$, displaying the validity of the KGW-formalism to model fermions at different regimes. Here we have used $E_0=1$.}
    \label{Klein_vs_Dirac}
\end{figure}
For the case with a time-dependent electric field, in \cref{Klein_vs_Dirac}, we plot the ratio $n_k/n_D$ versus $N$.
We can note that for small $N$, i.e., in the constant field limit, the ratio is very close to $0.5$ as expected. This is a good agreement between numerical and analytical solutions. For $N = 200$, we have $\hbar\omega=2m$ and it is thus possible to create an electron-positron pair from a single quanta. For this case, the relative rate of the KGW pair production has a minimum. Generally, the ratio $n_k/n_D$ tends to be smaller than $0.5$ when processes involving few quanta (as opposed to the pure Schwinger mechanism) are possible to occur. This is because the probability of producing pairs from few quanta is smaller in the KGW-description than in the DHW-description. Looking specifically at one-quanta processes, this fact is confirmed in the linear damping results (where the linear damping is due to single quanta pair creation) as given by \cref{ImD_KGW}-\cref{ImD_DHW}. For frequencies $\hbar \omega \approx 2m_e$, we have $p_{res}$ close to zero, and thus the created pairs have very small momentum. From \cref{ImD_KGW}, the damping of the wave in the KGW-formalism scales as $p_{res}^3$, compared to the DHW-formalism, where the wave damping is linear in $p_{res}$. For higher frequencies, we have larger $p_{res}$ and this lead to a somewhat larger value of $n_k/n_D$, but it is still well below $0.5$. The reason for the saturation of pair creation for high frequencies in the KGW-model is the scaling $\omega_i \sim \omega_r^{-4}$ in \cref{ImD_KGW}, more than compensating for the scaling $\omega_i \sim p_{res}^{3}$.

\begin{figure}
    \centering
    \includegraphics[width=\linewidth,height=10cm]{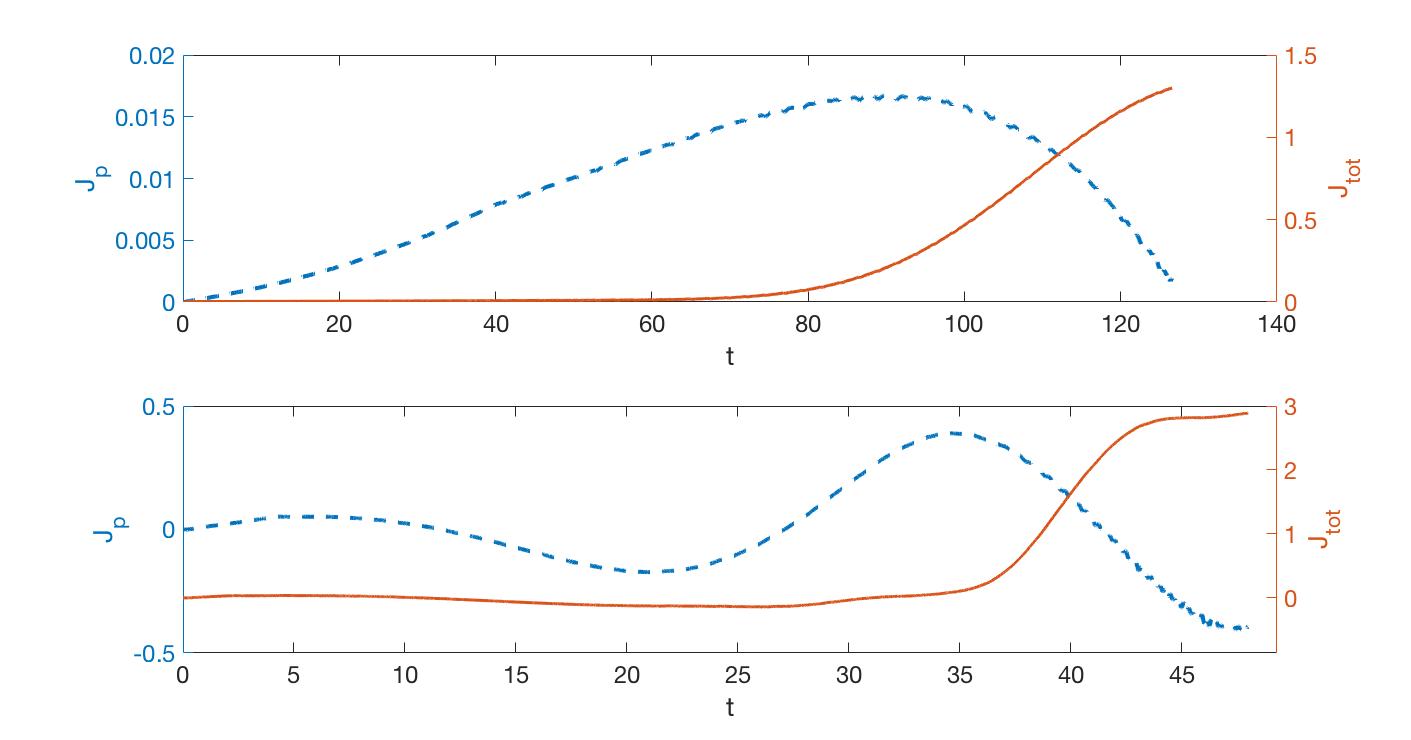}
    \caption{Polarization current $J_p$ and total current $J_{tot}$ plotted over time for $N=(2,20)$ in the upper and lower panels, respectively. Note the different scales for $J_p$ and $J_{tot}$}
    \label{Polarization}
\end{figure}

Since the KGW-formalism is spinless, only free currents are created. By contrast, the total current in the DHW-formalism is the sum of free current, the magnetization current \cite{pondspin1}, and the polarization current $J_p$ \cite{asenjo}. Note, however, that in the absence of spatial variations, the magnetization current vanishes. For pulses with $\omega \ll \omega_{ce}$, the polarization current is small due to $J_p=\partial P/\partial t$, and the slow (compared to the Compton frequency) temporal scale. Here $P$ is the polarization due to the particle spin properties. The smallness of the polarization current is confirmed in \cref{Polarization}, where the polarization and total currents are plotted over time by solving the equations of the DHW-formalism using the pulse in \cref{Pulse}. For the upper panel of \cref{Polarization} where we used $N=2$, i.e., $\omega=0.02\omega_{ce}$, the polarization current is roughly only around $1\%$ of the total current. When the frequency is increased to $N=20$, the polarization current increases to about $10\%$ of the total current. As the ratio $n_k/n_D=1/2$  comes from the difference in the vacuum contribution only, the deviation from this value depends on the different dynamics of the KGW and DHW descriptions. For the present case, as long as the polarization current density $J_p$ associated with the spin is small compared to the total current density, the dynamical influence of spin effects is comparatively small. Thus it is not surprising that the ratio $n_k/n_D=1/2$ is relatively unaffected by the spin dynamics. However, a careful comparison of \cref{Klein_vs_Dirac} and \cref{Polarization} shows that the deviation from $n_k/n_D=1/2$ coincides with an increase in the relative importance of the spin dynamics. 

\begin{figure}
    \centering
    \includegraphics[width=\linewidth,height=10cm]{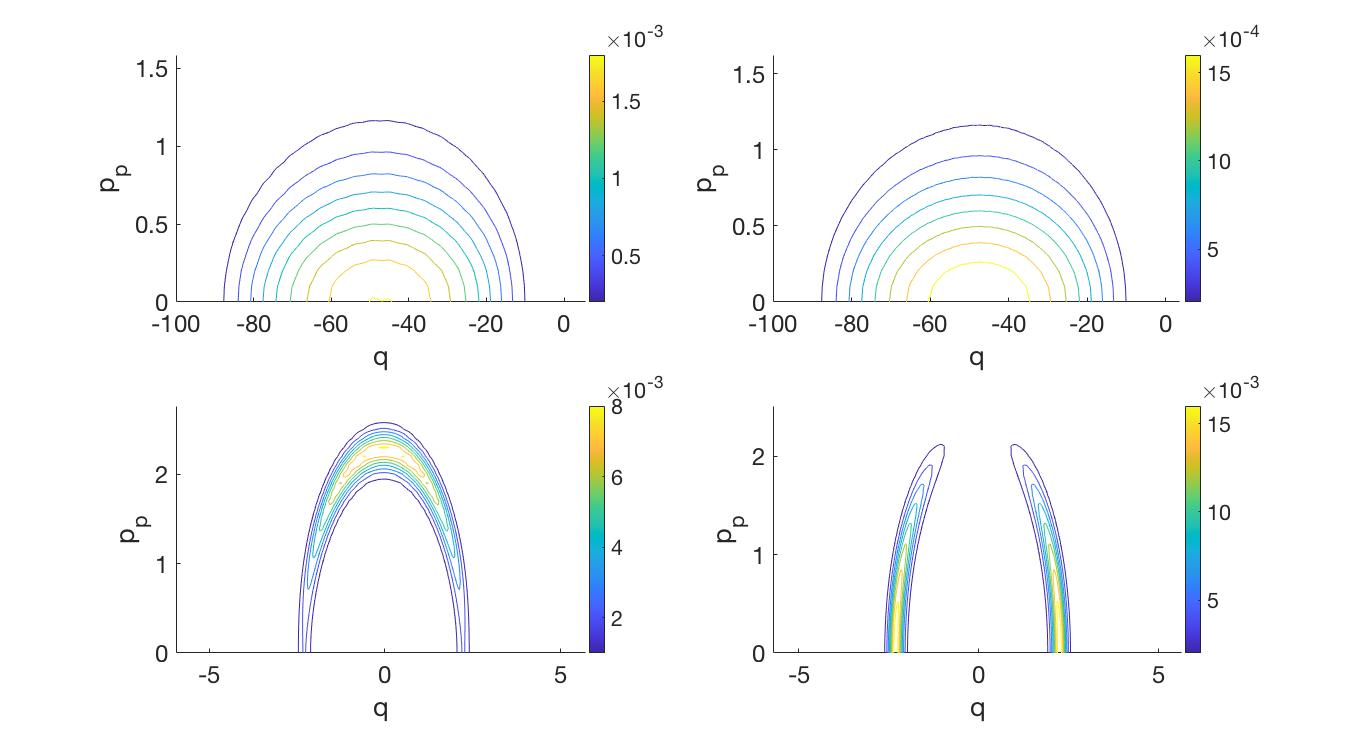}
    \caption{Color mappings of the momentum distribution of the created pairs from DHW $n_D(q,p_{\bot})$ (the first column) and KGW $n_k(q,p_{\bot})$ (the second column). In the first panel, we have used $\omega=0.02 \omega_{ce}$ , $\tau_0=\pi$, $b=20$ and $E_0=1$ and in the second panel we have $\omega=5\omega_{ce}$, $\tau_0=\pi$, $b=2$ and $E_0=2$. }
    \label{Contour}
\end{figure}

Next, we would like to make contact between the analytical and numerical calculations. Specifically, we will study where in momentum space the pairs are produced. In \cref{Contour}, contour curves over the pair momentum density, i.e. the integrand in \cref{Rate_KGW} is shown. The two figures in the upper panel show the contour for DHW-expression $n_D(p_{\bot},q)$ (the figure to the left) and the KGW-expression $n_k(p_{\bot},q)$ (the figure to the right) using the pulse in \cref{Pulse} with $\omega=0.02 \omega_{cr}$ , $\tau_0=\pi$, $b=20$ and $E_0=1$.
In the upper panel, we can see that created pairs have the same momentum distribution in both of the models. It should be emphasized that we here have used the toy model with the vacuum contribution twice as large as for a scalar field in the KGW calculations, to better mimic the behavior of fermions.

In the lower panel of \cref{Contour}, we have used the pulse in \cref{Pulse} with $\omega=5\omega_{ce}$, $\tau_0=\pi$, $b=2$ and $E_0=2$.
For such a high frequency, single quanta pair-creation processes dominate. Clearly, the polarization current $J_p$ is important in this regime.
As can be seen, in spite that pairs are created close to the region of resonant momenta, satisfying the resonance condition $m^2+p_{res}^2=\hbar^2\omega^2$,
the different models have different momentum distributions nevertheless. For the KGW-model (the figure to the right), the pairs created tend to have a higher value of $\abs{q}$, and a smaller value of $p_{\bot}$. For the DHW-model, the distribution in momentum space is the opposite.

To clarify the reason for this difference, we will rewrite the previous expressions \cref{Disp_hom_KGW,Disp_hom_DHW} in cylindrical momentum coordinates. In this case, the dispersion relation \cref{Disp_hom_KGW} for the KGW-case becomes

 \begin{multline}
 \label{Disp_hom_cyl}
    \omega^2=\frac{e^2}{4\pi^2 \hbar^3} \int dp_{\bot}p_{\bot}dp_z \frac{1/ \epsilon}{\hbar ^2\omega^2-4\epsilon^2} \times  \\
     \bigg[ 2f(\hbar ^2\omega^2-4\epsilon_{\perp}^2)
     +\hbar^2\omega^2 \frac{p_z^2}{\epsilon^2}
     \bigg]
 \end{multline}
and the dispersion relation for the DHW-case is written

\begin{multline}
\label{DHW_Disp}
    \omega^2= -\frac{e^2}{2\pi^2\hbar^3 }\int dp_{\bot}p_{\bot}dp_z \ \frac{ \epsilon_{\bot}^2/ \epsilon}{\hbar^2\omega^2-4\epsilon^2}\times \\
    \bigg(
f-  \frac{\hbar^2\omega^2}{4\epsilon^2}
    \bigg) 
\end{multline}
where $\epsilon_{\bot}=\sqrt{m^2+p_{\bot}^2}$. 
Note that we have regained $p_z$ rather than $q$ here \cite{Comment} and also keep the un-normalized variables of \cref{Section2}.
The key to understanding the lower panels of \cref{Contour} are the scalings with $p_z$ and $p_{\bot}$ of the integrands in \cref{Disp_hom_cyl,DHW_Disp}. What contributes to the damping rates is the pair creation that results from the pole contribution. That is, pair creation comes from an integration over resonant momenta $p_{res}$. The resulting pair creation will have a distribution in momentum space reflecting the magnitude of the integrand as a function of $p_z$ and $p_{\bot}$ along the curve given by a constant $p_{res}$.
As we are now considering pair-creation in vacuum rather than a plasma, we can inspect the integrand in \cref{Disp_hom_cyl} with only the vacuum contribution present, in which case the integrand reads

 \begin{equation}
    \int dp_{\bot}p_{\bot}dp_z \frac{p_z^2/ \epsilon^3}{\hbar ^2\omega^2-4\epsilon^2} 
 \end{equation}
Noting that the momenta which contributes to single-quanta pair creation will be close to the resonant momenta, we see that for resonant values with large $p_z$ ($q$) and small $p_{\perp}$, we will get the dominant pair creation contribution. This is in agreement with the lower right panel of \cref{Contour} as $p_{\bot} \sim 0$ gives a negligible pair creation rate.

For the left figure of the lower panel in \cref{Contour}, we have the contour curves of the momentum distribution for the created pairs in the DHW-formalism. As noted above, we see that the created pairs tend to have a high perpendicular momentum. This effect has been seen in previous works \cite{2021,Krajwska}. Moreover, it is further confirmed by considering only the vacuum contribution to the integrand in \cref{DHW_Disp}

\begin{equation}
    \int dp_{\bot}p_{\bot}dp_z \ \frac{ \epsilon_{\bot}^2/ \epsilon^3}{\hbar^2\omega^2-4\epsilon^2}
\end{equation}
We note that for resonant momenta, the integrand is largest for high perpendicular momentum and has a smaller value when $\abs{p_z}$ is large, explaining why the momentum distribution of the single-quanta pair creation in the DHW-formalism has the momentum distribution that we see in \cref{Contour}.  
\begin{figure}
    \centering
    \includegraphics[width=\linewidth, height=10 cm]{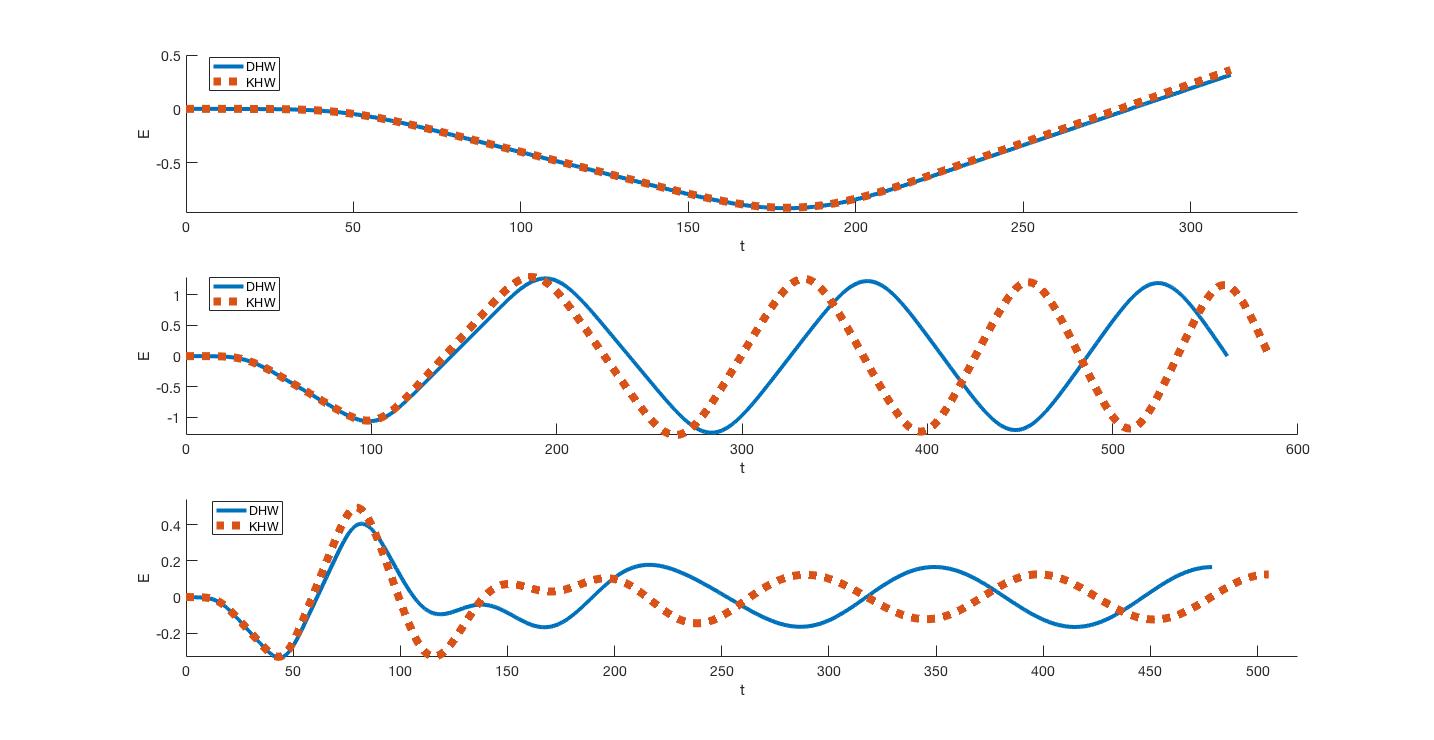}
    \caption{The time profile of the self-consistent electric field from the KGW and DHW-formalism for three values of the frequency of the initial electric pulse $N=(0.02,0.04,0.1)$. }
    \label{Time_profile}
\end{figure}

Next, we will include the effects due to back-reaction, i.e., we will solve \cref{Ampers_law_KGW} together with \cref{PDE-system_KGW} using the pulse in \cref{Pulse}.
We start from vacuum and divide the electric field into the self-consistent electric field, given by the current sources generated by the created pairs, and the external electric field which is prescribed and given by the same type of temporal profile as before.  

In \cref{Time_profile}, we have a plot of the self-consistent electric field (as the prescribed part is the same) over time comparing the results from DHW and KGW. In the first panel, we have used $N=2$, i.e., $\omega=0.02\omega_{ce}$. For this low frequency, we see a very good agreement between DHW and KGW. The good agreement is dependent on the choice to double the background vacuum contribution in the KGW-model, i.e., we have imitated initial conditions for the fermions, even though the dynamical evolution follows that of a scalar field, neglecting the spin.

In the second panel, we use a slightly higher frequency $N=4$. While the agreement is still good,  we see a graudally growing phase-shift between the DHW and KGW models. While the fraction $n_K/n_D$ was almost unchanged when the frequency was increased to this value, when the self-consistent dynamic is involved, we see a higher sensitivity to the frequency. Increasing the frequency even higher to $0.1\omega_{ce}$, the difference between DHW and KGW is even more pronounced. Here, we see a more dramatic phase-shift at an early stage and a qualitatively different evolution of the wave amplitude. Thus, modeling electron-positron pair creation with the KGW-model requires lower frequencies in the case of self-consistent dynamics. 

\subsection{Non-degeneracy effects}
\begin{figure}
    \centering
    \includegraphics[width=\linewidth,height=10cm]{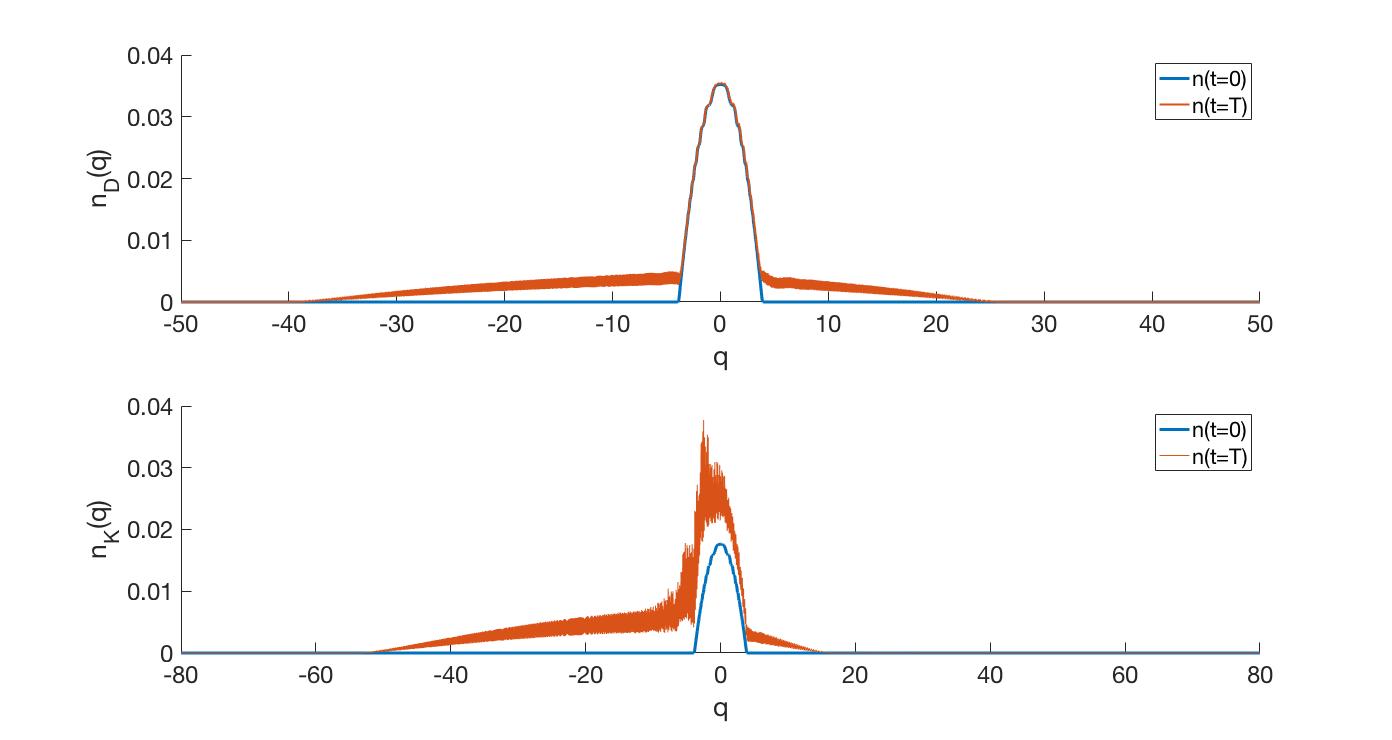}
    \caption{The number density $n_{D/K}(q)= \int dp_{\bot}p_{\bot}n_{D/K}(q,p_{\bot})$ as a function of the canonical momentum is plotted for $E_0=4$, $\mu=4$ and $T=0.02$ for DHW and KGW. Here, the solid curve is for n(t=0) and the .... is for n(t=T).}
    \label{Pauli_blocking}
\end{figure}

In this subsection, we would like to illustrate the mechanism of
Pauli blocking.  This mechanism is always present in the DHW
formalism, but not in the KGW-formalism as the model is spinless. However, whether or not Pauli blocking is an important feature
dynamically, depends on the number of free states where pairs
can be created. In particular, a degenerate initial distribution
will block the low-energy states in momentum space, which is
the region where particles are most easily created. Thus, in this subsection, we will start with a plasma initially and solve \cref{PDE-system_KGW,Ampers_law_KGW} and the corresponding equations in the DHW-formalism. As initial conditions in the runs, we let $E(t=0)=E_0$ and $A(t=0)=0$. The initial plasma is represented by a Fermi-Dirac distribution,  
\begin{equation}
\label{Fermi}
f(q,p_{\bot } )=\frac{1}{1+\exp{(\epsilon-\mu)/T}}
\end{equation}
where $\mu$ is the chemical potential and $T$ is the temperature (we normalize both $\mu$ and $T$ with the electron mass). For $\mu\gg T$, the chemical potential fulfills $\mu\approx E_F$, where $E_F$ is the Fermi energy, in which case the distribution will be degenerate (i.e., we will have $f \approx 1$, i.e. all electron states filled, for energies smaller than the Fermi energy, and $f \approx 0$ otherwise). Naturally, non-degenerate initial distributions are still possible when using \cref{Fermi}, by letting $\mu < T$. Also, by picking a large but negative chemical potential, we get a Maxwell-Boltzmann type of distribution.

In \cref{Pauli_blocking}, we demonstrate the Pauli blocking mechanism.
In the upper panel, we have considered the DHW-formalism using a degenerate distribution with $\mu=4$ and $T=0.02$, where all the low-energy electron states are filled, preventing further pair-creation in the filled region. As shown, due to a strong initial field $E_0=4$, pairs are still created at a high rate, but the low-energy region is perfectly blocked. This is seen in \cref{Pauli_blocking} by comparing the momentum distribution $n_D(q)$ after an oscillation period with the initial momentum distribution $n_D(t=0)$. Specifically, it should be noted that the two curves coincide (no further pair-creation) for a low or modest parallel momentum.
In the lower panel of \cref{Pauli_blocking}, we have used the KGW-formalism using the same input data as in the upper panel. We note that pairs are still created for low parallel momentum, in contrast to the DHW case. The effect is expected as the KGW model does not include the effect of the Pauli blocking {\it dynamically}, even if a degenerate distribution is chosen as an initial condition. 

\section{Conclusion}
\label{Section4}
In this paper, we have compared pair-creation in the KGW-formalism, based on a Wigner transform of the Klein-Gordon equation, and pair-creation in the DHW-formalism, based on a Wigner transform of the Dirac equation. Firstly, studying high-frequency pair creation in low amplitude linear theory, we have noted that while the DHW-formalism and the KGW-formalism share qualitative features (e.g. pair-creation when the same resonance condition is fulfilled), the expression for the wave damping rates due to pair creation are fundamentally different. 

Generally, the agreement between KGW and DHW is better for lower frequencies of the electromagnetic field. In particular, if one compensates for the larger magnitude of the vacuum background in the DHW-case (by picking a vacuum background a factor two larger than for a genuine scalar field), the KGW and DHW equations agree perfectly for the pair-production rate in the zero-frequency limit.

Increasing the frequency of the electric field, the agreement gradually deteriorates. The lack of agreement is correlated with a simultaneous growth of the relative contribution of the polarization current density to the total current density. As the polarization current is a direct consequence of the electron spin, this result is to be expected. 

Studying pair-production including the back-reaction from the self-consistent current density created by the pairs, the sensitivity to a finite frequency is increased. That is, we need a lower frequency of the electromagnetic field to still get a good agreement between the DHW-theory and the KGW-theory. Practically speaking $\omega\sim 0.04\omega_c$ or smaller,  Moreover, for a self-consistent theory it is crucial to magnify the vacuum background by a factor two in the KGW-case (as compared to a genuine scalar field), as the nonlinear dynamics otherwise will be heavily modified, in case the self-consistent response has the wrong magnitude. 

Furthermore, we have studied the dynamics for large plasma densities, where Pauli blocking is prominent. As expected, in the DHW-formalism pairs can not be created in a region of momentum space where the states are occupied. The same does not apply for the KGW-formalism, naturally.

The present study is motivated by a desire to use the KGW-formalism whenever it is a valid approximation, as the 16 scalar equations of the DHW-formalism (as compared to the 4 scalar equations of the KGW-formalism) are considerably more difficult to solve, both numerically and analytically. While the present study has certain restrictions, e.g. with only minor exceptions we have considered electrostatic fields and the homogeneous limit, we expect that many of the present findings apply in more general scenarios. In particular, we expect that the relatively good agreement between the DHW-formalism and the KGW-formalism for low frequencies generalizes to cases of spatial variations and electromagnetic fields, at least qualitatively. However, more studies are needed to give a definitive answer.

 \bibliography{References}

\begin{thebibliography}{40}%
\makeatletter
\providecommand \@ifxundefined [1]{%
 \@ifx{#1\undefined}
}%
\providecommand \@ifnum [1]{%
 \ifnum #1\expandafter \@firstoftwo
 \else \expandafter \@secondoftwo
 \fi
}%
\providecommand \@ifx [1]{%
 \ifx #1\expandafter \@firstoftwo
 \else \expandafter \@secondoftwo
 \fi
}%
\providecommand \natexlab [1]{#1}%
\providecommand \enquote  [1]{``#1''}%
\providecommand \bibnamefont  [1]{#1}%
\providecommand \bibfnamefont [1]{#1}%
\providecommand \citenamefont [1]{#1}%
\providecommand \href@noop [0]{\@secondoftwo}%
\providecommand \href [0]{\begingroup \@sanitize@url \@href}%
\providecommand \@href[1]{\@@startlink{#1}\@@href}%
\providecommand \@@href[1]{\endgroup#1\@@endlink}%
\providecommand \@sanitize@url [0]{\catcode `\\12\catcode `\$12\catcode
  `\&12\catcode `\#12\catcode `\^12\catcode `\_12\catcode `\%12\relax}%
\providecommand \@@startlink[1]{}%
\providecommand \@@endlink[0]{}%
\providecommand \url  [0]{\begingroup\@sanitize@url \@url }%
\providecommand \@url [1]{\endgroup\@href {#1}{\urlprefix }}%
\providecommand \urlprefix  [0]{URL }%
\providecommand \Eprint [0]{\href }%
\providecommand \doibase [0]{https://doi.org/}%
\providecommand \selectlanguage [0]{\@gobble}%
\providecommand \bibinfo  [0]{\@secondoftwo}%
\providecommand \bibfield  [0]{\@secondoftwo}%
\providecommand \translation [1]{[#1]}%
\providecommand \BibitemOpen [0]{}%
\providecommand \bibitemStop [0]{}%
\providecommand \bibitemNoStop [0]{.\EOS\space}%
\providecommand \EOS [0]{\spacefactor3000\relax}%
\providecommand \BibitemShut  [1]{\csname bibitem#1\endcsname}%
\let\auto@bib@innerbib\@empty
\bibitem [{\citenamefont {Danson}\ \emph {et~al.}(2019)\citenamefont {Danson},
  \citenamefont {Haefner}, \citenamefont {Bromage}, \citenamefont {Butcher},
  \citenamefont {Chanteloup}, \citenamefont {Chowdhury}, \citenamefont
  {Galvanauskas}, \citenamefont {Gizzi}, \citenamefont {Hein}, \citenamefont
  {Hillier} \emph {et~al.}}]{ELI}%
  \BibitemOpen
  \bibfield  {author} {\bibinfo {author} {\bibfnamefont {C.~N.}\ \bibnamefont
  {Danson}}, \bibinfo {author} {\bibfnamefont {C.}~\bibnamefont {Haefner}},
  \bibinfo {author} {\bibfnamefont {J.}~\bibnamefont {Bromage}}, \bibinfo
  {author} {\bibfnamefont {T.}~\bibnamefont {Butcher}}, \bibinfo {author}
  {\bibfnamefont {J.-C.~F.}\ \bibnamefont {Chanteloup}}, \bibinfo {author}
  {\bibfnamefont {E.~A.}\ \bibnamefont {Chowdhury}}, \bibinfo {author}
  {\bibfnamefont {A.}~\bibnamefont {Galvanauskas}}, \bibinfo {author}
  {\bibfnamefont {L.~A.}\ \bibnamefont {Gizzi}}, \bibinfo {author}
  {\bibfnamefont {J.}~\bibnamefont {Hein}}, \bibinfo {author} {\bibfnamefont
  {D.~I.}\ \bibnamefont {Hillier}}, \emph {et~al.},\ }\bibfield  {title}
  {\bibinfo {title} {Petawatt and exawatt class lasers worldwide},\ }\href@noop
  {} {\bibfield  {journal} {\bibinfo  {journal} {High Power Laser Science and
  Engineering}\ }\textbf {\bibinfo {volume} {7}},\ \bibinfo {pages} {e54}
  (\bibinfo {year} {2019})}\BibitemShut {NoStop}%
\bibitem [{\citenamefont {Cartlidge}(2018)}]{SEL}%
  \BibitemOpen
  \bibfield  {author} {\bibinfo {author} {\bibfnamefont {E.}~\bibnamefont
  {Cartlidge}},\ }\href@noop {} {\bibinfo {title} {The light fantastic}}
  (\bibinfo {year} {2018})\BibitemShut {NoStop}%
\bibitem [{\citenamefont {Di~Piazza}\ \emph {et~al.}(2012)\citenamefont
  {Di~Piazza}, \citenamefont {M\"uller}, \citenamefont {Hatsagortsyan},\ and\
  \citenamefont {Keitel}}]{Review1}%
  \BibitemOpen
  \bibfield  {author} {\bibinfo {author} {\bibfnamefont {A.}~\bibnamefont
  {Di~Piazza}}, \bibinfo {author} {\bibfnamefont {C.}~\bibnamefont {M\"uller}},
  \bibinfo {author} {\bibfnamefont {K.~Z.}\ \bibnamefont {Hatsagortsyan}},\
  and\ \bibinfo {author} {\bibfnamefont {C.~H.}\ \bibnamefont {Keitel}},\
  }\bibfield  {title} {\bibinfo {title} {Extremely high-intensity laser
  interactions with fundamental quantum systems},\ }\href
  {https://doi.org/10.1103/RevModPhys.84.1177} {\bibfield  {journal} {\bibinfo
  {journal} {Rev. Mod. Phys.}\ }\textbf {\bibinfo {volume} {84}},\ \bibinfo
  {pages} {1177} (\bibinfo {year} {2012})}\BibitemShut {NoStop}%
\bibitem [{\citenamefont {Fedotov}\ \emph {et~al.}(2023)\citenamefont
  {Fedotov}, \citenamefont {Ilderton}, \citenamefont {Karbstein}, \citenamefont
  {King}, \citenamefont {Seipt}, \citenamefont {Taya},\ and\ \citenamefont
  {Torgrimsson}}]{Anton}%
  \BibitemOpen
  \bibfield  {author} {\bibinfo {author} {\bibfnamefont {A.}~\bibnamefont
  {Fedotov}}, \bibinfo {author} {\bibfnamefont {A.}~\bibnamefont {Ilderton}},
  \bibinfo {author} {\bibfnamefont {F.}~\bibnamefont {Karbstein}}, \bibinfo
  {author} {\bibfnamefont {B.}~\bibnamefont {King}}, \bibinfo {author}
  {\bibfnamefont {D.}~\bibnamefont {Seipt}}, \bibinfo {author} {\bibfnamefont
  {H.}~\bibnamefont {Taya}},\ and\ \bibinfo {author} {\bibfnamefont
  {G.}~\bibnamefont {Torgrimsson}},\ }\bibfield  {title} {\bibinfo {title}
  {Advances in qed with intense background fields},\ }\href
  {https://doi.org/https://doi.org/10.1016/j.physrep.2023.01.003} {\bibfield
  {journal} {\bibinfo  {journal} {Physics Reports}\ }\textbf {\bibinfo {volume}
  {1010}},\ \bibinfo {pages} {1} (\bibinfo {year} {2023})}\BibitemShut
  {NoStop}%
\bibitem [{\citenamefont {Sauter}(1931)}]{Sauter}%
  \BibitemOpen
  \bibfield  {author} {\bibinfo {author} {\bibfnamefont {F.}~\bibnamefont
  {Sauter}},\ }\bibfield  {title} {\bibinfo {title} {{\"U}ber das verhalten
  eines elektrons im homogenen elektrischen feld nach der relativistischen
  theorie diracs},\ }\href@noop {} {\bibfield  {journal} {\bibinfo  {journal}
  {Zeitschrift f{\"u}r Physik}\ }\textbf {\bibinfo {volume} {69}},\ \bibinfo
  {pages} {742} (\bibinfo {year} {1931})}\BibitemShut {NoStop}%
\bibitem [{\citenamefont {Schwinger}(1951)}]{Schwinger}%
  \BibitemOpen
  \bibfield  {author} {\bibinfo {author} {\bibfnamefont {J.}~\bibnamefont
  {Schwinger}},\ }\bibfield  {title} {\bibinfo {title} {On gauge invariance and
  vacuum polarization},\ }\href {https://doi.org/10.1103/PhysRev.82.664}
  {\bibfield  {journal} {\bibinfo  {journal} {Phys. Rev.}\ }\textbf {\bibinfo
  {volume} {82}},\ \bibinfo {pages} {664} (\bibinfo {year} {1951})}\BibitemShut
  {NoStop}%
\bibitem [{\citenamefont {Hebenstreit}\ \emph {et~al.}(2010)\citenamefont
  {Hebenstreit}, \citenamefont {Alkofer},\ and\ \citenamefont {Gies}}]{Gies2}%
  \BibitemOpen
  \bibfield  {author} {\bibinfo {author} {\bibfnamefont {F.}~\bibnamefont
  {Hebenstreit}}, \bibinfo {author} {\bibfnamefont {R.}~\bibnamefont
  {Alkofer}},\ and\ \bibinfo {author} {\bibfnamefont {H.}~\bibnamefont
  {Gies}},\ }\bibfield  {title} {\bibinfo {title} {Schwinger pair production in
  space- and time-dependent electric fields: Relating the wigner formalism to
  quantum kinetic theory},\ }\href {https://doi.org/10.1103/PhysRevD.82.105026}
  {\bibfield  {journal} {\bibinfo  {journal} {Phys. Rev. D}\ }\textbf {\bibinfo
  {volume} {82}},\ \bibinfo {pages} {105026} (\bibinfo {year}
  {2010})}\BibitemShut {NoStop}%
\bibitem [{\citenamefont {Hebenstreit}\ \emph {et~al.}(2011)\citenamefont
  {Hebenstreit}, \citenamefont {Alkofer},\ and\ \citenamefont {Gies}}]{Gies1}%
  \BibitemOpen
  \bibfield  {author} {\bibinfo {author} {\bibfnamefont {F.}~\bibnamefont
  {Hebenstreit}}, \bibinfo {author} {\bibfnamefont {R.}~\bibnamefont
  {Alkofer}},\ and\ \bibinfo {author} {\bibfnamefont {H.}~\bibnamefont
  {Gies}},\ }\bibfield  {title} {\bibinfo {title} {Particle self-bunching in
  the schwinger effect in spacetime-dependent electric fields},\ }\href
  {https://doi.org/10.1103/PhysRevLett.107.180403} {\bibfield  {journal}
  {\bibinfo  {journal} {Phys. Rev. Lett.}\ }\textbf {\bibinfo {volume} {107}},\
  \bibinfo {pages} {180403} (\bibinfo {year} {2011})}\BibitemShut {NoStop}%
\bibitem [{\citenamefont {Aleksandrov}\ and\ \citenamefont
  {Kohlf\"urst}(2020)}]{Kohlfurst_spat1}%
  \BibitemOpen
  \bibfield  {author} {\bibinfo {author} {\bibfnamefont {I.~A.}\ \bibnamefont
  {Aleksandrov}}\ and\ \bibinfo {author} {\bibfnamefont {C.}~\bibnamefont
  {Kohlf\"urst}},\ }\bibfield  {title} {\bibinfo {title} {Pair production in
  temporally and spatially oscillating fields},\ }\href
  {https://doi.org/10.1103/PhysRevD.101.096009} {\bibfield  {journal} {\bibinfo
   {journal} {Phys. Rev. D}\ }\textbf {\bibinfo {volume} {101}},\ \bibinfo
  {pages} {096009} (\bibinfo {year} {2020})}\BibitemShut {NoStop}%
\bibitem [{\citenamefont {Kohlf\"urst}(2020)}]{Kohlfurst_spat2}%
  \BibitemOpen
  \bibfield  {author} {\bibinfo {author} {\bibfnamefont {C.}~\bibnamefont
  {Kohlf\"urst}},\ }\bibfield  {title} {\bibinfo {title} {Effect of
  time-dependent inhomogeneous magnetic fields on the particle momentum
  spectrum in electron-positron pair production},\ }\href
  {https://doi.org/10.1103/PhysRevD.101.096003} {\bibfield  {journal} {\bibinfo
   {journal} {Phys. Rev. D}\ }\textbf {\bibinfo {volume} {101}},\ \bibinfo
  {pages} {096003} (\bibinfo {year} {2020})}\BibitemShut {NoStop}%
\bibitem [{\citenamefont {Bulanov}\ \emph {et~al.}(2010)\citenamefont
  {Bulanov}, \citenamefont {Mur}, \citenamefont {Narozhny}, \citenamefont
  {Nees},\ and\ \citenamefont {Popov}}]{Colliding2}%
  \BibitemOpen
  \bibfield  {author} {\bibinfo {author} {\bibfnamefont {S.~S.}\ \bibnamefont
  {Bulanov}}, \bibinfo {author} {\bibfnamefont {V.~D.}\ \bibnamefont {Mur}},
  \bibinfo {author} {\bibfnamefont {N.~B.}\ \bibnamefont {Narozhny}}, \bibinfo
  {author} {\bibfnamefont {J.}~\bibnamefont {Nees}},\ and\ \bibinfo {author}
  {\bibfnamefont {V.~S.}\ \bibnamefont {Popov}},\ }\bibfield  {title} {\bibinfo
  {title} {Multiple colliding electromagnetic pulses: A way to lower the
  threshold of ${e}^{+}{e}^{\ensuremath{-}}$ pair production from vacuum},\
  }\href {https://doi.org/10.1103/PhysRevLett.104.220404} {\bibfield  {journal}
  {\bibinfo  {journal} {Phys. Rev. Lett.}\ }\textbf {\bibinfo {volume} {104}},\
  \bibinfo {pages} {220404} (\bibinfo {year} {2010})}\BibitemShut {NoStop}%
\bibitem [{\citenamefont {Gonoskov}\ \emph {et~al.}(2013)\citenamefont
  {Gonoskov}, \citenamefont {Gonoskov}, \citenamefont {Harvey}, \citenamefont
  {Ilderton}, \citenamefont {Kim}, \citenamefont {Marklund}, \citenamefont
  {Mourou},\ and\ \citenamefont {Sergeev}}]{Colliding1}%
  \BibitemOpen
  \bibfield  {author} {\bibinfo {author} {\bibfnamefont {A.}~\bibnamefont
  {Gonoskov}}, \bibinfo {author} {\bibfnamefont {I.}~\bibnamefont {Gonoskov}},
  \bibinfo {author} {\bibfnamefont {C.}~\bibnamefont {Harvey}}, \bibinfo
  {author} {\bibfnamefont {A.}~\bibnamefont {Ilderton}}, \bibinfo {author}
  {\bibfnamefont {A.}~\bibnamefont {Kim}}, \bibinfo {author} {\bibfnamefont
  {M.}~\bibnamefont {Marklund}}, \bibinfo {author} {\bibfnamefont
  {G.}~\bibnamefont {Mourou}},\ and\ \bibinfo {author} {\bibfnamefont
  {A.}~\bibnamefont {Sergeev}},\ }\bibfield  {title} {\bibinfo {title} {Probing
  nonperturbative qed with optimally focused laser pulses},\ }\href
  {https://doi.org/10.1103/PhysRevLett.111.060404} {\bibfield  {journal}
  {\bibinfo  {journal} {Phys. Rev. Lett.}\ }\textbf {\bibinfo {volume} {111}},\
  \bibinfo {pages} {060404} (\bibinfo {year} {2013})}\BibitemShut {NoStop}%
\bibitem [{\citenamefont {Torgrimsson}\ \emph {et~al.}(2018)\citenamefont
  {Torgrimsson}, \citenamefont {Schneider},\ and\ \citenamefont
  {Sch\"utzhold}}]{Colliding4}%
  \BibitemOpen
  \bibfield  {author} {\bibinfo {author} {\bibfnamefont {G.}~\bibnamefont
  {Torgrimsson}}, \bibinfo {author} {\bibfnamefont {C.}~\bibnamefont
  {Schneider}},\ and\ \bibinfo {author} {\bibfnamefont {R.}~\bibnamefont
  {Sch\"utzhold}},\ }\bibfield  {title} {\bibinfo {title} {Sauter-schwinger
  pair creation dynamically assisted by a plane wave},\ }\href
  {https://doi.org/10.1103/PhysRevD.97.096004} {\bibfield  {journal} {\bibinfo
  {journal} {Phys. Rev. D}\ }\textbf {\bibinfo {volume} {97}},\ \bibinfo
  {pages} {096004} (\bibinfo {year} {2018})}\BibitemShut {NoStop}%
\bibitem [{\citenamefont {Kohlf\"urst}\ \emph {et~al.}(2022)\citenamefont
  {Kohlf\"urst}, \citenamefont {Ahmadiniaz}, \citenamefont {Oertel},\ and\
  \citenamefont {Sch\"utzhold}}]{colliding3}%
  \BibitemOpen
  \bibfield  {author} {\bibinfo {author} {\bibfnamefont {C.}~\bibnamefont
  {Kohlf\"urst}}, \bibinfo {author} {\bibfnamefont {N.}~\bibnamefont
  {Ahmadiniaz}}, \bibinfo {author} {\bibfnamefont {J.}~\bibnamefont {Oertel}},\
  and\ \bibinfo {author} {\bibfnamefont {R.}~\bibnamefont {Sch\"utzhold}},\
  }\bibfield  {title} {\bibinfo {title} {Sauter-schwinger effect for colliding
  laser pulses},\ }\href {https://doi.org/10.1103/PhysRevLett.129.241801}
  {\bibfield  {journal} {\bibinfo  {journal} {Phys. Rev. Lett.}\ }\textbf
  {\bibinfo {volume} {129}},\ \bibinfo {pages} {241801} (\bibinfo {year}
  {2022})}\BibitemShut {NoStop}%
\bibitem [{\citenamefont {Narozhnyi}\ and\ \citenamefont
  {Nikishov}(1973)}]{ScalarQED7}%
  \BibitemOpen
  \bibfield  {author} {\bibinfo {author} {\bibfnamefont {N.}~\bibnamefont
  {Narozhnyi}}\ and\ \bibinfo {author} {\bibfnamefont {A.}~\bibnamefont
  {Nikishov}},\ }\bibfield  {title} {\bibinfo {title} {Pair production by},\
  }\href@noop {} {\bibfield  {journal} {\bibinfo  {journal} {Zh. Eksp. Teor.
  Fiz}\ }\textbf {\bibinfo {volume} {65}},\ \bibinfo {pages} {862} (\bibinfo
  {year} {1973})}\BibitemShut {NoStop}%
\bibitem [{\citenamefont {Kohlf\"urst}(2019)}]{Kohlfurst2}%
  \BibitemOpen
  \bibfield  {author} {\bibinfo {author} {\bibfnamefont {C.}~\bibnamefont
  {Kohlf\"urst}},\ }\bibfield  {title} {\bibinfo {title} {Spin states in
  multiphoton pair production for circularly polarized light},\ }\href
  {https://doi.org/10.1103/PhysRevD.99.096017} {\bibfield  {journal} {\bibinfo
  {journal} {Phys. Rev. D}\ }\textbf {\bibinfo {volume} {99}},\ \bibinfo
  {pages} {096017} (\bibinfo {year} {2019})}\BibitemShut {NoStop}%
\bibitem [{\citenamefont {Kohlf{\"u}rst}(2022)}]{Kohlfurst1}%
  \BibitemOpen
  \bibfield  {author} {\bibinfo {author} {\bibfnamefont {C.}~\bibnamefont
  {Kohlf{\"u}rst}},\ }\bibfield  {title} {\bibinfo {title} {The
  heisenberg-wigner formalism for transverse fields},\ }\href@noop {}
  {\bibfield  {journal} {\bibinfo  {journal} {arXiv preprint arXiv:2212.06057}\
  } (\bibinfo {year} {2022})}\BibitemShut {NoStop}%
\bibitem [{\citenamefont {Bloch}\ \emph {et~al.}(1999)\citenamefont {Bloch},
  \citenamefont {Mizerny}, \citenamefont {Prozorkevich}, \citenamefont
  {Roberts}, \citenamefont {Schmidt}, \citenamefont {Smolyansky},\ and\
  \citenamefont {Vinnik}}]{Bloch}%
  \BibitemOpen
  \bibfield  {author} {\bibinfo {author} {\bibfnamefont {J.~C.~R.}\
  \bibnamefont {Bloch}}, \bibinfo {author} {\bibfnamefont {V.~A.}\ \bibnamefont
  {Mizerny}}, \bibinfo {author} {\bibfnamefont {A.~V.}\ \bibnamefont
  {Prozorkevich}}, \bibinfo {author} {\bibfnamefont {C.~D.}\ \bibnamefont
  {Roberts}}, \bibinfo {author} {\bibfnamefont {S.~M.}\ \bibnamefont
  {Schmidt}}, \bibinfo {author} {\bibfnamefont {S.~A.}\ \bibnamefont
  {Smolyansky}},\ and\ \bibinfo {author} {\bibfnamefont {D.~V.}\ \bibnamefont
  {Vinnik}},\ }\bibfield  {title} {\bibinfo {title} {Pair creation: Back
  reactions and damping},\ }\href {https://doi.org/10.1103/PhysRevD.60.116011}
  {\bibfield  {journal} {\bibinfo  {journal} {Phys. Rev. D}\ }\textbf {\bibinfo
  {volume} {60}},\ \bibinfo {pages} {116011} (\bibinfo {year}
  {1999})}\BibitemShut {NoStop}%
\bibitem [{\citenamefont {Cooper}\ and\ \citenamefont
  {Mottola}(1989)}]{ScalarQED1}%
  \BibitemOpen
  \bibfield  {author} {\bibinfo {author} {\bibfnamefont {F.}~\bibnamefont
  {Cooper}}\ and\ \bibinfo {author} {\bibfnamefont {E.}~\bibnamefont
  {Mottola}},\ }\bibfield  {title} {\bibinfo {title} {Quantum back reaction in
  scalar qed as an initial-value problem},\ }\href
  {https://doi.org/10.1103/PhysRevD.40.456} {\bibfield  {journal} {\bibinfo
  {journal} {Phys. Rev. D}\ }\textbf {\bibinfo {volume} {40}},\ \bibinfo
  {pages} {456} (\bibinfo {year} {1989})}\BibitemShut {NoStop}%
\bibitem [{\citenamefont {Kluger}\ \emph {et~al.}(1991)\citenamefont {Kluger},
  \citenamefont {Eisenberg}, \citenamefont {Svetitsky}, \citenamefont
  {Cooper},\ and\ \citenamefont {Mottola}}]{ScalarQED3}%
  \BibitemOpen
  \bibfield  {author} {\bibinfo {author} {\bibfnamefont {Y.}~\bibnamefont
  {Kluger}}, \bibinfo {author} {\bibfnamefont {J.}~\bibnamefont {Eisenberg}},
  \bibinfo {author} {\bibfnamefont {B.}~\bibnamefont {Svetitsky}}, \bibinfo
  {author} {\bibfnamefont {F.}~\bibnamefont {Cooper}},\ and\ \bibinfo {author}
  {\bibfnamefont {E.}~\bibnamefont {Mottola}},\ }\bibfield  {title} {\bibinfo
  {title} {Pair production in a strong electric field},\ }\href@noop {}
  {\bibfield  {journal} {\bibinfo  {journal} {Physical Review Letters}\
  }\textbf {\bibinfo {volume} {67}},\ \bibinfo {pages} {2427} (\bibinfo {year}
  {1991})}\BibitemShut {NoStop}%
\bibitem [{\citenamefont {Kim}\ and\ \citenamefont {Page}(2006)}]{ScalarQED4}%
  \BibitemOpen
  \bibfield  {author} {\bibinfo {author} {\bibfnamefont {S.~P.}\ \bibnamefont
  {Kim}}\ and\ \bibinfo {author} {\bibfnamefont {D.~N.}\ \bibnamefont {Page}},\
  }\bibfield  {title} {\bibinfo {title} {Schwinger pair production in electric
  and magnetic fields},\ }\href@noop {} {\bibfield  {journal} {\bibinfo
  {journal} {Physical Review D}\ }\textbf {\bibinfo {volume} {73}},\ \bibinfo
  {pages} {065020} (\bibinfo {year} {2006})}\BibitemShut {NoStop}%
\bibitem [{\citenamefont {Shi}\ \emph {et~al.}(2018)\citenamefont {Shi},
  \citenamefont {Xiao}, \citenamefont {Qin},\ and\ \citenamefont
  {Fisch}}]{ScalarQED2}%
  \BibitemOpen
  \bibfield  {author} {\bibinfo {author} {\bibfnamefont {Y.}~\bibnamefont
  {Shi}}, \bibinfo {author} {\bibfnamefont {J.}~\bibnamefont {Xiao}}, \bibinfo
  {author} {\bibfnamefont {H.}~\bibnamefont {Qin}},\ and\ \bibinfo {author}
  {\bibfnamefont {N.~J.}\ \bibnamefont {Fisch}},\ }\bibfield  {title} {\bibinfo
  {title} {Simulations of relativistic quantum plasmas using real-time lattice
  scalar qed},\ }\href {https://doi.org/10.1103/PhysRevE.97.053206} {\bibfield
  {journal} {\bibinfo  {journal} {Phys. Rev. E}\ }\textbf {\bibinfo {volume}
  {97}},\ \bibinfo {pages} {053206} (\bibinfo {year} {2018})}\BibitemShut
  {NoStop}%
\bibitem [{\citenamefont {Bialynicki-Birula}\ \emph {et~al.}(1991)\citenamefont
  {Bialynicki-Birula}, \citenamefont {Gornicki},\ and\ \citenamefont
  {Rafelski}}]{Birula}%
  \BibitemOpen
  \bibfield  {author} {\bibinfo {author} {\bibfnamefont {I.}~\bibnamefont
  {Bialynicki-Birula}}, \bibinfo {author} {\bibfnamefont {P.}~\bibnamefont
  {Gornicki}},\ and\ \bibinfo {author} {\bibfnamefont {J.}~\bibnamefont
  {Rafelski}},\ }\bibfield  {title} {\bibinfo {title} {Phase-space structure of
  the dirac vacuum},\ }\href@noop {} {\bibfield  {journal} {\bibinfo  {journal}
  {Physical Review D}\ }\textbf {\bibinfo {volume} {44}},\ \bibinfo {pages}
  {1825} (\bibinfo {year} {1991})}\BibitemShut {NoStop}%
\bibitem [{\citenamefont {Best}\ \emph {et~al.}(1993)\citenamefont {Best},
  \citenamefont {Gornicki},\ and\ \citenamefont {Greiner}}]{Best}%
  \BibitemOpen
  \bibfield  {author} {\bibinfo {author} {\bibfnamefont {C.}~\bibnamefont
  {Best}}, \bibinfo {author} {\bibfnamefont {P.}~\bibnamefont {Gornicki}},\
  and\ \bibinfo {author} {\bibfnamefont {W.}~\bibnamefont {Greiner}},\
  }\bibfield  {title} {\bibinfo {title} {The phase-space structure of the
  klein-gordon field},\ }\href@noop {} {\bibfield  {journal} {\bibinfo
  {journal} {Annals of Physics}\ }\textbf {\bibinfo {volume} {225}},\ \bibinfo
  {pages} {169} (\bibinfo {year} {1993})}\BibitemShut {NoStop}%
\bibitem [{\citenamefont {Li}\ \emph {et~al.}(2010)\citenamefont {Li},
  \citenamefont {Wang}, \citenamefont {Dulat},\ and\ \citenamefont
  {Ma}}]{KG_wigner_Review}%
  \BibitemOpen
  \bibfield  {author} {\bibinfo {author} {\bibfnamefont {K.}~\bibnamefont
  {Li}}, \bibinfo {author} {\bibfnamefont {J.}~\bibnamefont {Wang}}, \bibinfo
  {author} {\bibfnamefont {S.}~\bibnamefont {Dulat}},\ and\ \bibinfo {author}
  {\bibfnamefont {K.}~\bibnamefont {Ma}},\ }\bibfield  {title} {\bibinfo
  {title} {Wigner functions for klein-gordon oscillators in non-commutative
  space},\ }\href@noop {} {\bibfield  {journal} {\bibinfo  {journal}
  {International Journal of Theoretical Physics}\ }\textbf {\bibinfo {volume}
  {49}},\ \bibinfo {pages} {134} (\bibinfo {year} {2010})}\BibitemShut
  {NoStop}%
\bibitem [{\citenamefont {Santos}\ and\ \citenamefont
  {Silva}(2005)}]{KG_wigner}%
  \BibitemOpen
  \bibfield  {author} {\bibinfo {author} {\bibfnamefont {J.}~\bibnamefont
  {Santos}}\ and\ \bibinfo {author} {\bibfnamefont {L.}~\bibnamefont {Silva}},\
  }\bibfield  {title} {\bibinfo {title} {Wigner-moyal description of free
  variable mass klein-gordon fields},\ }\href@noop {} {\bibfield  {journal}
  {\bibinfo  {journal} {Journal of mathematical physics}\ }\textbf {\bibinfo
  {volume} {46}},\ \bibinfo {pages} {102901} (\bibinfo {year}
  {2005})}\BibitemShut {NoStop}%
\bibitem [{\citenamefont {Wong}(2010)}]{KG_wigner_hydrodyn}%
  \BibitemOpen
  \bibfield  {author} {\bibinfo {author} {\bibfnamefont {C.-Y.}\ \bibnamefont
  {Wong}},\ }\bibfield  {title} {\bibinfo {title} {Klein--gordon equation in
  hydrodynamical form},\ }\href@noop {} {\bibfield  {journal} {\bibinfo
  {journal} {Journal of mathematical physics}\ }\textbf {\bibinfo {volume}
  {51}},\ \bibinfo {pages} {122304} (\bibinfo {year} {2010})}\BibitemShut
  {NoStop}%
\bibitem [{\citenamefont {Zhuang}\ and\ \citenamefont
  {Heinz}(1998)}]{ScalarQED5}%
  \BibitemOpen
  \bibfield  {author} {\bibinfo {author} {\bibfnamefont {P.}~\bibnamefont
  {Zhuang}}\ and\ \bibinfo {author} {\bibfnamefont {U.}~\bibnamefont {Heinz}},\
  }\bibfield  {title} {\bibinfo {title} {Equal-time hierarchies for quantum
  transport theory},\ }\href {https://doi.org/10.1103/PhysRevD.57.6525}
  {\bibfield  {journal} {\bibinfo  {journal} {Phys. Rev. D}\ }\textbf {\bibinfo
  {volume} {57}},\ \bibinfo {pages} {6525} (\bibinfo {year}
  {1998})}\BibitemShut {NoStop}%
\bibitem [{\citenamefont {Li}\ \emph {et~al.}(2019)\citenamefont {Li},
  \citenamefont {Xie},\ and\ \citenamefont {Li}}]{ScalarQED6}%
  \BibitemOpen
  \bibfield  {author} {\bibinfo {author} {\bibfnamefont {Z.~L.}\ \bibnamefont
  {Li}}, \bibinfo {author} {\bibfnamefont {B.~S.}\ \bibnamefont {Xie}},\ and\
  \bibinfo {author} {\bibfnamefont {Y.~J.}\ \bibnamefont {Li}},\ }\bibfield
  {title} {\bibinfo {title} {Boson pair production in arbitrarily polarized
  electric fields},\ }\href {https://doi.org/10.1103/PhysRevD.100.076018}
  {\bibfield  {journal} {\bibinfo  {journal} {Phys. Rev. D}\ }\textbf {\bibinfo
  {volume} {100}},\ \bibinfo {pages} {076018} (\bibinfo {year}
  {2019})}\BibitemShut {NoStop}%
\bibitem [{\citenamefont {Nikishov}(1970)}]{Nikishov}%
  \BibitemOpen
  \bibfield  {author} {\bibinfo {author} {\bibfnamefont {A.}~\bibnamefont
  {Nikishov}},\ }\bibfield  {title} {\bibinfo {title} {Pair production by a
  constant external field},\ }\href@noop {} {\bibfield  {journal} {\bibinfo
  {journal} {Sov. Phys. JETP}\ }\textbf {\bibinfo {volume} {30}},\ \bibinfo
  {pages} {660} (\bibinfo {year} {1970})}\BibitemShut {NoStop}%
\bibitem [{\citenamefont {Feshbach}\ and\ \citenamefont
  {Villars}(1958)}]{Feshbach}%
  \BibitemOpen
  \bibfield  {author} {\bibinfo {author} {\bibfnamefont {H.}~\bibnamefont
  {Feshbach}}\ and\ \bibinfo {author} {\bibfnamefont {F.}~\bibnamefont
  {Villars}},\ }\bibfield  {title} {\bibinfo {title} {Elementary relativistic
  wave mechanics of spin 0 and spin 1/2 particles},\ }\href@noop {} {\bibfield
  {journal} {\bibinfo  {journal} {Reviews of Modern Physics}\ }\textbf
  {\bibinfo {volume} {30}},\ \bibinfo {pages} {24} (\bibinfo {year}
  {1958})}\BibitemShut {NoStop}%
\bibitem [{\citenamefont {Al-Naseri}\ \emph {et~al.}(2021)\citenamefont
  {Al-Naseri}, \citenamefont {Zamanian},\ and\ \citenamefont {Brodin}}]{2021}%
  \BibitemOpen
  \bibfield  {author} {\bibinfo {author} {\bibfnamefont {H.}~\bibnamefont
  {Al-Naseri}}, \bibinfo {author} {\bibfnamefont {J.}~\bibnamefont
  {Zamanian}},\ and\ \bibinfo {author} {\bibfnamefont {G.}~\bibnamefont
  {Brodin}},\ }\bibfield  {title} {\bibinfo {title} {Plasma dynamics and vacuum
  pair creation using the dirac-heisenberg-wigner formalism},\ }\href
  {https://doi.org/10.1103/PhysRevE.104.015207} {\bibfield  {journal} {\bibinfo
   {journal} {Phys. Rev. E}\ }\textbf {\bibinfo {volume} {104}},\ \bibinfo
  {pages} {015207} (\bibinfo {year} {2021})}\BibitemShut {NoStop}%
\bibitem [{\citenamefont {Brodin}\ and\ \citenamefont
  {Zamanian}(2022)}]{RMPP-review}%
  \BibitemOpen
  \bibfield  {author} {\bibinfo {author} {\bibfnamefont {G.}~\bibnamefont
  {Brodin}}\ and\ \bibinfo {author} {\bibfnamefont {J.}~\bibnamefont
  {Zamanian}},\ }\bibfield  {title} {\bibinfo {title} {Quantum kinetic theory
  of plasmas},\ }\href@noop {} {\bibfield  {journal} {\bibinfo  {journal}
  {Reviews of Modern Plasma Physics}\ }\textbf {\bibinfo {volume} {6}},\
  \bibinfo {pages} {4} (\bibinfo {year} {2022})}\BibitemShut {NoStop}%
\bibitem [{\citenamefont {Al-Naseri}\ and\ \citenamefont
  {Brodin}(2022)}]{2022linear}%
  \BibitemOpen
  \bibfield  {author} {\bibinfo {author} {\bibfnamefont {H.}~\bibnamefont
  {Al-Naseri}}\ and\ \bibinfo {author} {\bibfnamefont {G.}~\bibnamefont
  {Brodin}},\ }\bibfield  {title} {\bibinfo {title} {Linear pair-creation
  damping of high-frequency plasma oscillation},\ }\href@noop {} {\bibfield
  {journal} {\bibinfo  {journal} {Physics of Plasmas}\ }\textbf {\bibinfo
  {volume} {29}},\ \bibinfo {pages} {042106} (\bibinfo {year}
  {2022})}\BibitemShut {NoStop}%
\bibitem [{\citenamefont {Brodin}\ \emph {et~al.}(2023)\citenamefont {Brodin},
  \citenamefont {Al-Naseri}, \citenamefont {Zamanian}, \citenamefont
  {Torgrimsson},\ and\ \citenamefont {Eliasson}}]{PRESchwinger}%
  \BibitemOpen
  \bibfield  {author} {\bibinfo {author} {\bibfnamefont {G.}~\bibnamefont
  {Brodin}}, \bibinfo {author} {\bibfnamefont {H.}~\bibnamefont {Al-Naseri}},
  \bibinfo {author} {\bibfnamefont {J.}~\bibnamefont {Zamanian}}, \bibinfo
  {author} {\bibfnamefont {G.}~\bibnamefont {Torgrimsson}},\ and\ \bibinfo
  {author} {\bibfnamefont {B.}~\bibnamefont {Eliasson}},\ }\bibfield  {title}
  {\bibinfo {title} {Plasma dynamics at the schwinger limit and beyond},\
  }\href {https://doi.org/10.1103/PhysRevE.107.035204} {\bibfield  {journal}
  {\bibinfo  {journal} {Phys. Rev. E}\ }\textbf {\bibinfo {volume} {107}},\
  \bibinfo {pages} {035204} (\bibinfo {year} {2023})}\BibitemShut {NoStop}%
\bibitem [{\citenamefont {Press}\ \emph {et~al.}(2007)\citenamefont {Press},
  \citenamefont {Teukolsky}, \citenamefont {Vetterling},\ and\ \citenamefont
  {Flannery}}]{Leapfrog}%
  \BibitemOpen
  \bibfield  {author} {\bibinfo {author} {\bibfnamefont {W.~H.}\ \bibnamefont
  {Press}}, \bibinfo {author} {\bibfnamefont {S.~A.}\ \bibnamefont
  {Teukolsky}}, \bibinfo {author} {\bibfnamefont {W.~T.}\ \bibnamefont
  {Vetterling}},\ and\ \bibinfo {author} {\bibfnamefont {B.~P.}\ \bibnamefont
  {Flannery}},\ }\href@noop {} {\emph {\bibinfo {title} {Numerical recipes 3rd
  edition: The art of scientific computing}}}\ (\bibinfo  {publisher}
  {Cambridge university press},\ \bibinfo {year} {2007})\BibitemShut {NoStop}%
\bibitem [{\citenamefont {Stefan}\ \emph {et~al.}(2011)\citenamefont {Stefan},
  \citenamefont {Zamanian}, \citenamefont {Brodin}, \citenamefont {Misra},\
  and\ \citenamefont {Marklund}}]{pondspin1}%
  \BibitemOpen
  \bibfield  {author} {\bibinfo {author} {\bibfnamefont {M.}~\bibnamefont
  {Stefan}}, \bibinfo {author} {\bibfnamefont {J.}~\bibnamefont {Zamanian}},
  \bibinfo {author} {\bibfnamefont {G.}~\bibnamefont {Brodin}}, \bibinfo
  {author} {\bibfnamefont {A.~P.}\ \bibnamefont {Misra}},\ and\ \bibinfo
  {author} {\bibfnamefont {M.}~\bibnamefont {Marklund}},\ }\bibfield  {title}
  {\bibinfo {title} {Ponderomotive force due to the intrinsic spin in extended
  fluid and kinetic models},\ }\href@noop {} {\bibfield  {journal} {\bibinfo
  {journal} {Physical Review E}\ }\textbf {\bibinfo {volume} {83}},\ \bibinfo
  {pages} {036410} (\bibinfo {year} {2011})}\BibitemShut {NoStop}%
\bibitem [{\citenamefont {Asenjo}\ \emph {et~al.}(2012)\citenamefont {Asenjo},
  \citenamefont {Zamanian}, \citenamefont {Marklund}, \citenamefont {Brodin},\
  and\ \citenamefont {Johansson}}]{asenjo}%
  \BibitemOpen
  \bibfield  {author} {\bibinfo {author} {\bibfnamefont {F.~A.}\ \bibnamefont
  {Asenjo}}, \bibinfo {author} {\bibfnamefont {J.}~\bibnamefont {Zamanian}},
  \bibinfo {author} {\bibfnamefont {M.}~\bibnamefont {Marklund}}, \bibinfo
  {author} {\bibfnamefont {G.}~\bibnamefont {Brodin}},\ and\ \bibinfo {author}
  {\bibfnamefont {P.}~\bibnamefont {Johansson}},\ }\bibfield  {title} {\bibinfo
  {title} {Semi-relativistic effects in spin-1/2 quantum plasmas},\ }\href@noop
  {} {\bibfield  {journal} {\bibinfo  {journal} {New Journal of Physics}\
  }\textbf {\bibinfo {volume} {14}},\ \bibinfo {pages} {073042} (\bibinfo
  {year} {2012})}\BibitemShut {NoStop}%
\bibitem [{Com()}]{Comment}%
  \BibitemOpen
  \href@noop {} {\bibinfo  {journal} {We note that the difference between $q$
  and $p_z$ is not important for high frequency pair creation as this process
  will take place also for small/modest electric field, which leads to a small
  $A$}\ }\BibitemShut {NoStop}%
\bibitem [{\citenamefont {Krajewska}\ and\ \citenamefont
  {Kami\ifmmode~\acute{n}\else \'{n}\fi{}ski}(2019)}]{Krajwska}%
  \BibitemOpen
\bibfield  {journal} {  }\bibfield  {author} {\bibinfo {author} {\bibfnamefont
  {K.}~\bibnamefont {Krajewska}}\ and\ \bibinfo {author} {\bibfnamefont
  {J.~Z.}\ \bibnamefont {Kami\ifmmode~\acute{n}\else \'{n}\fi{}ski}},\
  }\bibfield  {title} {\bibinfo {title} {Threshold effects in electron-positron
  pair creation from the vacuum: Stabilization and longitudinal versus
  transverse momentum sharing},\ }\href
  {https://doi.org/10.1103/PhysRevA.100.012104} {\bibfield  {journal} {\bibinfo
   {journal} {Phys. Rev. A}\ }\textbf {\bibinfo {volume} {100}},\ \bibinfo
  {pages} {012104} (\bibinfo {year} {2019})}\BibitemShut {NoStop}%
\end{thebibliography}%


\end{document}